\documentclass[11pt,a4paper,english]{article}

\usepackage{graphicx}
\usepackage{amssymb,latexsym}
\usepackage{amsmath}
\usepackage{epsf}
\usepackage{pdfsync}
\usepackage{cite}
\usepackage{epsfig}
\usepackage[parfill]{parskip}
\usepackage[matrix,arrow,color]{xy}
\usepackage[usenames]{color}
\usepackage{hyperref}
\usepackage{bbm}
\usepackage{dsfont}
\usepackage{stmaryrd}

\input{xy}
\xyoption{all}

\input{epsf}

\makeatletter

\catcode`@=11
\renewcommand{\section}
{\@startsection{section}{1}{0pt}{\medskipamount}{\medskipamount}{\large\bf}}
\makeatletter\renewcommand{\subsection}
{\@startsection{subsection}{2}{\z@}{-3.25ex plus -1ex minus -.2ex}
{1.5ex plus .2ex}{\it }}

\numberwithin{equation}{section}
\catcode`@=12

\newcommand{\ba}{\begin{eqnarray*}}
\newcommand{\ea}{\end{eqnarray*}}
\newcommand{\ban}{\begin{eqnarray}}
\newcommand{\ean}{\end{eqnarray}}

\newcommand{\zbar}{\bar{z}}

\newcommand{\cW}{{\cal W}}

\newcommand{\cM}{{\cal M}}

\newcommand{\cL}{{\cal L}}

\newcommand{\mbf}[1]{{\boldsymbol {#1} }}

\def\e{{\,\rm e}\,}
\newcommand{\wt}{\widetilde}

\def\ii{{\,{\rm i}\,}}
\def\dd{{\rm d}}

\def\beq{\begin{equation}}
\def\bee{\begin{equation}}
\def\eeq{\end{equation}}
\def\bea{\begin{eqnarray}}
\def\eea{\end{eqnarray}}
\def\bd{\begin{displaymath}}
\def\ed{\end{displaymath}}

\newcommand{\Cint}{\int\kern-10.5pt-\kern7pt}

\newcommand{\be}{\begin{equation}}
\newcommand{\ee}{\end{equation}}

\newcommand\fverbit{\egroup\item[\fbox{\unhbox\pippobox}]}
\newbox\pippobox

{
\def\d{\delta}

\def\pa{\partial}

\def\bs{\bar{\sigma}}

\def\w{\wedge}

\def\rd{\overset{\shortrightarrow}{\partial}}
\def\ldelta{\overset{\shortleftarrow}{\delta}}
\def\rdelta{\overset{\shortrightarrow}{\delta}}

\def\bs#1{\boldsymbol{#1}}
\def\bv#1#2{\left( {#1},{#2} \right)_{\rm{BV}} }
\def\rm#1{\mathrm{#1}}
\def\wt#1{\widetilde{#1}}

\def\be{\begin{equation}}
\def\ee{\end{equation}}
\def\bea{\begin{eqnarray}}
\def\eea{\end{eqnarray}}

\begin{document}

\begin{titlepage}
\setcounter{page}{0}
\begin{flushright}
\small
EMPG--18--03
\end{flushright}

\vskip 1.8cm

\begin{center}

{\Large\bf AKSZ constructions for topological membranes on $\mbf{G_2}$-manifolds}

\vspace{15mm}

{\large\bf Zolt\'an K\"ok\'enyesi$^{(a),(b)}$, Annam{\'a}ria Sinkovics$^{(a)}$ and
 Richard~J.~Szabo$^{(c)}$}
\\[6mm]
\noindent{\em $^{(a)}$ Institute of Theoretical Physics\\ MTA-ELTE
  Theoretical Research Group \\ E\"otv\"os Lor\'and University \\ P\'azm\'any s. 1/A, 1117
  Budapest, Hungary} \\ Email: \ {\tt
  kokenyesiz@caesar.elte.hu} \ , \ {\tt
  sinkovics@general.elte.hu}\\[4mm]
\noindent{\em $^{(b)}$ Institute for Particle and Nuclear Physics\\ Wigner Research Center for
Physics\\
Konkoly-Thege Mikl{\'o}s {\'u}t 29-33, 1121 Budapest, Hungary} \\[4mm]
\noindent{\em $^{(c)}$ Department of Mathematics\\ Heriot-Watt
  University\\
Colin Maclaurin Building, Riccarton, Edinburgh EH14 4AS, UK\\ 
Maxwell Institute for Mathematical Sciences, Edinburgh, UK\\
The Higgs Centre for Theoretical Physics, Edinburgh, UK}\\
Email: \ {\tt
  R.J.Szabo@hw.ac.uk}

\vspace{30mm}

\begin{abstract}
\noindent
We consider AKSZ constructions of BV actions for closed topological
membranes, and their dimensional reductions to topological string
sigma-models. Two inequivalent AKSZ constructions for topological
membranes on $G_2$-manifolds are proposed, in each of which the two existing
topological membrane theories appear as different gauge fixed
versions. Their dimensional reductions give new AKSZ constructions for
the topological A-model, which on further dimensional reduction gives
an AKSZ formulation of supersymmetric quantum mechanics. We show that
the two AKSZ membrane models originate through worldvolume dimensional
reduction of a single AKSZ threebrane
theory, which gives the
standard 2-Courant bracket as the underlying derived bracket. Double
dimensional reduction of the twisted topological threebrane theory on a
circle yields the standard Courant sigma-model for string theory with NS--NS flux.
\end{abstract}

\end{center}


\end{titlepage}



{\baselineskip=18pt
\tableofcontents
}

\newpage

\section{Introduction}

Topological M-theory was originally proposed as a unification of the topological A- and B-models~\cite{Dijkgraaf2004,Nekrasov2004}, and is intended to capture a topological sector of physical M-theory. It can be constructed on seven-dimensional manifolds of $G_2$-holonomy where it has reduced $\mathcal{N}=1$ supersymmetry. The theory of~\cite{Dijkgraaf2004} is based on a Hitchin-type form theory of $G_2$-manifolds, and its dimensional reduction on a circle gives Hitchin's form theories of the topological A- and B-models. 

The A- and B-models have worldsheet formulations as string theories where they are given by two-dimensional topological sigma-models. Hence it is natural to expect that topological M-theory has a worldvolume formulation and its fundamental objects are topological membranes. Two different membrane theories have been proposed for this purpose. One is constructed using the Mathai-Quillen formalism in~\cite{MQAni2005}, which reduces on a circle to the Mathai-Quillen construction of the A-model~\cite{MQWu1995,MQBlau1995} and its path integral localizes on associative three-cycles. The other one is constructed in~\cite{Bonelli2005b} as a BRST gauge fixed version of the simple topological action constructed by pullback to the membrane worldvolume of the harmonic three-form associated to the $G_2$-structure, which also reduces to the A-model and localizes on associative three-cycles. 

Both types of topological membranes are intended to be the fundamental objects of the same theory, which inevitably raises the question of whether they can be described within a single membrane model. In this paper we aim to give a unified treatment of these objects by describing some new aspects of the AKSZ formulation for topological membranes and topological string sigma-models.
AKSZ formulations provide natural geometric methods for constructing BV quantized sigma-models which circumvent the difficulties involved in finding the BV extension of a classical action with degenerate symmetries~\cite{AKSZ1997,Cattaneo2001,Bouwknegt:2011vn,Ikeda2012}. They produce examples of topological field theories of Schwarz-type in arbitrary dimensionality which include well-known cases such as Chern-Simons theory, BF-theory and the Poisson sigma-model. Restrictions to special gauge fixing action functionals also yield examples of topological field theories of Witten-type, including the A/B-models and
Rozansky-Witten theory. 
In this paper we propose two different BV quantized sigma-models for topological membranes on $G_2$-manifolds given by the AKSZ formulation, which each give back the membrane theories discussed above in particular gauges. Our distinct AKSZ membrane theories have the special feature that they can be unified within a single AKSZ threebrane sigma-model, in which the derived bracket is the same as the anomaly-free current algebra of topological membranes induced on the generalized tangent bundle ${T} \oplus \bigwedge^2 {T}^*$ of $G_2$-manifolds~\cite{Bonelli2005a}.

One of the main motivations for studying AKSZ constructions is in the context of fluxes and generalized geometry in string theory and M-theory. Fluxes in compactifications of type~II string theory appear as twists of the Courant algebroid structure of the T-duality inspired generalized tangent bundle~\cite{Hull2005,Shelton2005,Blumenhagen2012,Heller2016}. Courant algebroids are in one-to-one correspondence to three-dimensional topological AKSZ sigma-models with target QP-manifold of degree 2, which are called Courant sigma-models~\cite{Roytenberg2002,Park2000,Ikeda2003,Roytenberg2002b,Hofman2002,Hofman2002a,Roytenberg2007}. 
Courant sigma-models geometrize fluxes in the sense that they are uplifts of string sigma-models to one higher
dimension which can accomodate fluxes~\cite{Bouwknegt:2011vn,Ikeda2012,Richard2012,Chatzistavrakidis2015}. 

The topological A- and B-models can also be described by gauge fixed
AKSZ sigma-models. They have been extensively studied by introducing AKSZ
membrane models with generalized complex structures arising from
generalized geometry, which reduce to the A- and
B-models~\cite{Zucchini2004,Zucchini2005,Pestun2006,Ikeda2007,Stojevic2005}. In
forthcoming work~\cite{Kokenyesi2018InProg} we will show that the
A-model is closely related to the contravariant Courant sigma-model
of~\cite{Bessho2015}.

Inspired by this analogy, in this paper we construct AKSZ membrane sigma-models which dimensionally reduce to give new AKSZ constructions for the A-model after gauge fixing and canonical transformation. Performing a further dimensional reduction of one of these string models then gives a novel AKSZ construction for supersymmetric quantum mechanics. We also propose an AKSZ topological three-brane theory, which reproduces our membrane sigma-models through a worldvolume dimensional reduction, and yields the standard 2-Courant bracket as its derived bracket, which fits it into the context of exceptional generalized geometry in M-theory. We also study the three-brane theory with a four-form flux twisting, and show that upon double dimensional reduction on a circle it reproduces the standard $H$-flux twisted Courant sigma-model.

This paper is organized as follows. In \S\ref{sec:AKSZBackground} we give a relatively detailed overview of various salient features of the AKSZ construction, together with a few examples of relevance for this paper such as the Poisson, Courant and 2-Courant sigma-models, and we derive the correspondence between flux twistings of these topological field theories. In \S\ref{sec:AKSZmembrG2} we introduce two AKSZ constructions for each of the topological membrane theories of~\cite{MQAni2005} and~\cite{Bonelli2005b}, and describe their underlying derived bracket algebra along with their origins as dimensional reductions of AKSZ topological threebrane theories. In \S\ref{sec:AKSZtopAmod} we calculate their dimensional reductions and show that the reduced AKSZ sigma-models are AKSZ constructions for the topological A-model, which can be similarly unified through worldvolume dimensional reduction of a single Courant sigma-model. We dimensionally reduce it further in \S\ref{sec:SQM} and get an AKSZ construction for supersymmetric quantum mechanics. Finally, we close with some concluding remarks and outlook on further applications of our constructions in~\S\ref{sec:Conc}. 

\section{Aspects of the AKSZ construction}
\label{sec:AKSZBackground}

In this section we survey some pertinent background about the AKSZ
construction and BV quantization, and describe several relevant
examples that we will encounter throughout this paper.

\subsection{AKSZ sigma-models}
\label{sec:AKSZingeneral}

We begin by briefly introducing the ingredients of AKSZ theory. A more complete review can be found in \cite{Ikeda2012}.
The AKSZ construction is a BV quantized sigma-model formulation, and it gives a geometric solution to the classical master equation 
\be
\bv{\bs{S}}{\bs{S}}=0
\ee
given by the BV bracket, which imposes BRST symmetry. The solution $\mbf S$ is called the AKSZ action, which is just a BV action.

Two classes of supermanifolds enter the AKSZ formalism. The `source' $(\cW,Q_\cW,\mu)$ consists of a differential graded (dg-)manifold, i.e. a graded manifold $\cW$ equiped with a cohomological vector field\footnote{A vector field
$Q$ is cohomological if it is of degree~1 and its Lie derivative $\cL_Q$
squares
to zero.} $Q_\cW$, and a measure $\mu$ which is invariant under
$Q_\cW$. The `target' $(\cM,Q_\gamma,\omega)$ is a symplectic
dg-manifold, i.e. a graded manifold $\cM$ with a cohomological vector
field $Q_\gamma$, and a graded symplectic form $\omega$ for which
$Q_\gamma$ is a Hamiltonian vector field:\footnote{The
  operation $\iota_Q$ denotes contraction of a differential form along the vector field $Q$.}  $\iota_{Q_\gamma}\omega=\dd
\gamma$ for some Hamiltonian function $\gamma$ on $\cM$.

In this paper we are interested in the construction of $d$-dimensional
topological sigma-models for closed branes. Hence we take
$\cW=T[1]\Sigma_{d}$, the tangent bundle of a $d$-dimensional closed
and oriented worldvolume manifold $\Sigma_{d}$ with the degree of its
fibers shifted by~1, which is isomorphic to the exterior algebra of
differential forms $(\Omega(\Sigma_{d}),\wedge)$. We choose the
cohomological vector field $Q_\cW$ corresponding to the de Rham
differential, which in local affine coordinates $\hat z^{\hat\mu}=(\sigma^\mu,\theta^\mu)\in T[1]\Sigma_{d}$, with degree~0 coordinates $\sigma^\mu$ on $\Sigma_d$ and degree~1 fiber coordinates $\theta^\mu$, has the form\footnote{Repeated upper and lower indices are always implicitly understood to be summed over.} $Q_\cW=\theta^\mu\, \frac\partial{\partial\sigma^\mu}=:\mbf D$.
 The measure in local coordinates can be written in the form
 $\mu=\dd^d\hat z:= \dd^{d} \sigma \ \dd^{d} \theta$. 

The AKSZ space of fields is a mapping space
\be
\mbf\cM = {\sf Map}\big(T[1]\Sigma_d\,,\,\cM\big)
\ee
consisting of smooth maps from $(\cW,Q_\cW,\mu)$ to $(\cM,Q_\gamma,\omega)$. In order to reproduce the BV formalism, the symplectic structure $\omega$ is taken to be of degree $d-1$, so that the Hamiltonian function $\gamma$ is of degree $d$. We can introduce local coordinates on $\mbf\cM$ via the superfields
\be
\hat{\mbf X}{}^{\hat\imath}(\hat z^{\hat\mu}) \, = \, \mbf\phi^*(\hat
X^{\hat\imath}\,)(\hat z^{\hat\mu}) \  ,
\ee
for local coordinates $\hat z^{\hat\mu}\in \cW$, $\hat
X^{\hat\imath}\in \cM$ and $\mbf\phi\in\mbf\cM$. Then the de~Rham
differential on $\mbf\cM$ is given for $\cW=T[1]\Sigma_{d}$ by the vector field
\be
\mbf\delta = (-1)^d\ \int_{T[1]\Sigma_{d}}\, \dd^d \hat z \ \mbf\delta \hat{\mbf X}{}^{\hat\imath}(\hat{z})\, \frac\rdelta{\delta\hat{\mbf X}{}^{\hat\imath}(\hat{z})}
\ee
with ghost number 1. 
The cohomological vector fields $Q_\cW$ and $Q_\gamma$ induce a
cohomological vector field $\bs{Q}$ on $\mbf\cM$ in the following
way. For $\mbf\phi\in \mbf\cM$ and $\hat z\in \cW$, define
\bea
(\bs{Q}_0 \, \mbf\phi)(\hat z)  = \dd \mbf\phi (\hat z) \, Q_\cW (\hat
z) \qquad \mbox{and} \qquad
(\bs{Q}_\gamma \, \mbf\phi)(\hat z) = Q_\gamma\big(\mbf\phi(\hat z)\big) \ .
\eea
Then $\mbf\cM$ is a dg-manifold with cohomological vector field
\be
\bs{Q}  \, = \,  \bs{Q}_0 + \bs{Q}_\gamma \ . 
\ee

Given an $n$-form $\alpha\in\Omega^n(\cM)$, we can lift it to an $n$-form $\bs{\alpha}\in\Omega^n(\mbf\cM)$ by transgression to the mapping space as
\be
\bs{\alpha} \, = \, \int_\cW\, \mu\ \mathrm{ev}^* (\alpha) \ ,
\ee
where $\mathrm{ev}\,:\, \cW \times \mbf\cM \, \rightarrow \cM $ is the
evaluation map. This definition allows us to think of $\bs{\alpha}$ as
an $n$-form functional of the fields in $\mbf\cM$, on which the de Rham differential acts as
\be
\mbf\delta \mbf\alpha \, = \, \mbf\delta \int_\cW\, \mu\ \mathrm{ev}^* (\alpha) \, = \, \int_\cW\, \mu\ \mathrm{ev}^* (\dd\alpha) \ .
\ee
Due to the integration,
$\bs{\alpha}$ has ghost number $U(\alpha)-d$, where $U(\alpha)$
denotes the internal degree of $\alpha$. In particular, since
transgression is a chain map, from the symplectic form $\omega$ on $\cM$
and a Liouville potential $\vartheta$, such that
$\omega=\dd\vartheta$, we get the symplectic form $\bs{\omega}$ of
degree $-1$ and Liouville potential $\bs{\vartheta}$ on $\mbf\cM$, such
that $\mbf\omega=\mbf\delta\mbf\vartheta$. Furthermore, the
cohomological vector field $\bs{Q}$ on $\mbf\cM$ is also Hamiltonian with
Hamiltonian function $-\iota_{\mbf Q_0}\mbf\vartheta + \bs{\gamma}$ of degree~0:
$\iota_{\bs{Q}_\gamma}\bs{\omega}= \bs{\delta} \bs{\gamma}$. In other words,
the mapping space of superfields $\mbf\cM$ is itself a symplectic dg-manifold.

The BV bracket $(\,\cdot\,,\,\cdot\,)_{\mathrm{BV}}$ is the graded Poisson bracket of degree~1 on $\mbf\cM$ defined from $\mbf\omega$, and it corresponds to the graded Poisson bracket $\{\,\cdot\,,\,\cdot\,\}$ of degree $-d+1$ on $\cM$ defined from $\omega$, since the transgression map $\int_\cW\,\mu\ \mathrm{ev}^*$ is a Lie algebra homomorphism from $(\cM,\{\,\cdot\,,\,\cdot\,\})$ to $(\mbf\cM,\bv{\,\cdot\,}{\,\cdot\,})$:  
\be 
\int_\cW\, \mu\ \mathrm{ev}^* \big(\{F,G\}\big) = \Big(\int_\cW\,\mu\ \mathrm{ev}^* (F)\,,\,\int_\cW\, \mu\ \mathrm{ev}^* (G)\Big)_{\mathrm{BV}} \  ,
\ee
where $F$ and $G$ are any local functions on $\cM$. In particular, the
cohomological vector fields can be represented through derived
brackets as
\be
Q_\gamma=\{\gamma,\,\cdot\,\} \qquad \mbox{and} \qquad \mbf Q=\mbf D + (\mbf\gamma,\,\cdot\,)_{\mathrm{BV}} \ ,
\ee
and the cocycle conditions $Q_\gamma^2=0$ and $\mbf Q^2=0$ are equivalent to $\{\gamma,\gamma\}=0$ and $(\mbf\gamma,\mbf\gamma)_{\mathrm{BV}}=0$.

It is now evident how to construct the desired BV action. For this, we choose a Liouville potential $\vartheta$ on $\cM$, and write the solution of the classical master equation, i.e.~the degree~0 AKSZ action $\mbf S$, on $\mbf\cM$ in the form
\be
\bs{S} \, = \, \bs{S}_0 \, + \, \bs{\gamma} \ ,
\ee 
where $\bs{S}_0=-\iota_{\bs{Q}_0}\bs\vartheta$ is the kinetic term and
the Hamiltonian function $\mbf\gamma$ on $\mbf\cM$ is the interaction
term. For a source superworldvolume $\cW=T[1]\Sigma_d$, one has explicitly
\be 
\mbf S = \int_{T[1]\Sigma_{d}}\,\dd^d\hat z\ \big(-\iota_{Q_\cW}\mathrm{ev}^*(\vartheta)+\mathrm{ev}^*(\gamma) \big) \ .
\ee
In the BV formalism, the cohomological vector field $\mbf Q$
corresponds to the BRST charge which generates BRST transformations of
superfields and the BRST cohomology of the mapping space~$\mbf\cM$.

\medskip

{\underline{\sl Canonical transformations.} \ }  
In the BV formalism, the phase space $\mbf\cM$ of superfields is only
defined modulo canonical transformations. These are the maps of
$\mbf\cM$ that leave the BV bracket structure invariant. A canonical transformation is associated to a degree $d-1$ function
$\alpha$ on $\cM$. We use the notation $\delta_\alpha:=\{\, \cdot \,,\alpha\}$ for the
corresponding Hamiltonian vector field, and 
$\e^{\delta_\alpha}$ and $\e^{\delta_{\bs{\alpha}}}$ for the
respective canonical transformations given by the adjoint actions of
the respective Poisson brackets. The action of the canonical
transformation on $\bs{\gamma}$ is given by
$\e^{\delta_{\bs{\alpha}}}\bs{\gamma}=\int_\cW\,\mu\ \mathrm{ev}^*
(\e^{\delta_\alpha} \gamma)$, with $\e^{\delta_\alpha}\gamma=\gamma+Q_\gamma\alpha+O(\alpha^2)$, which preserves the classical master
equation as
\be
\big\{\e^{\delta_\alpha}\gamma,\e^{\delta_\alpha}\gamma\big\}
=\e^{\delta_\alpha}\{\gamma,\gamma\} = 0 \ ,
\ee
due to $\{\gamma,\gamma\}=0$. Then the AKSZ action $\bs{S}_0 \, + \, \bs{\gamma}$ is equivalent to $\bs{S}_0 \, + \, \e^{\delta_{\bs{\alpha}}}\bs{\gamma}$ up to a canonical transformation, which at first order shifts the action by a BRST-exact term.

A duality transformation in the AKSZ formalism is defined as a symplectomorphism $\bs{f}$, which is a diffeomorphism between underlying symplectic manifolds 
\be
\bs{f} \, : \, (\mbf\cM,\bs{\omega}) \, \longrightarrow \, (\mbf\cM',\bs{\omega}'\,) \  ,
\ee
satisfying
\be
\bs{f}^* \bs{\omega}' = \bs{\omega} \  .
\ee
In other words, $\mbf f$ is a coordinate transformation on symplectic manifolds which leaves the symplectic structure invariant. Then the canonical transformation $\e^{\delta_{\bs{\alpha}}}$ is a duality transformation as well.

\medskip

{\underline{\sl QP-manifolds.} \ }  
A common choice of target for the AKSZ construction is to take $\cM$
to be an N-manifold, which is a graded manifold with no coordinates of
negative degree. In this case the triple $(\cM,Q_\gamma,\omega)$ is
called a QP-manifold of degree $n=d-1$; if the N-manifold $\cM$ is
concentrated in degrees $0,1,\dots,n$, then $(\cM,Q_\gamma,\omega)$ is
called a symplectic Lie $n$-algebroid, and it arises from an $n$-graded vector bundle over the degree~0 body $M=\cM_0$ of $\cM$~\cite{Ikeda2012}; in particular, functions of degree $n-1$ can be identified with sections of a vector bundle $E\to M$ equiped with the structure of a Leibniz algebroid. For example, in the simplest dimension $d=1$ with target a degree~0 QP-manifold, one necessarily has $Q_\gamma=0$ and thus a symplectic Lie 0-algebroid is just an ordinary symplectic manifold $(\cM,\omega)$; in this case the degree~1 Hamiltonian function $\gamma$ is locally constant on $\cM$ and the AKSZ construction produces a topological quantum mechanics given as a one-dimensional Chern-Simons theory whose Chern-Simons form is a Liouville potential $\vartheta$~\cite{Gwilliam2011,GradyLi2015}.

In the following we will describe the AKSZ topological field theories
associated with the first few non-trivial members in the hierarchy of
QP-structures on the target manifold for dimensions $d=2,3,4$, in the
context of the string and membrane models of interest in this
paper. Later on we shall also deal with targets that have negative
degree coordinates and hence unravel new constructions even in low dimension.

\subsection{Gauge fixing in the superfield formalism}
\label{sec:gaugefixing}

The entire field content of a system with degenerate symmetries is
usually specified by separating it into `fields', which includes the
original physical and
ghost fields from the BRST picture, and dual `antifields', which correspond
to the equations of motion and define canonically conjugate variables
with respect to the symplectic phase space structure on the space
$\mbf\cM$ of all fields.
In AKSZ constructions the fields and antifields are not distinguished
from the onset. The theory is specified once the antifields are assigned, and different choices yield different field theories. 

In the usual BV quantized theories, the fields and antifields are
distinguished from the start. One chooses a gauge fixing fermion
$\Psi[\phi]$, which is a functional of the fields $\phi^a$
(but not the antifields) of ghost number $U=-1$, and then the
antifields $\phi_a^+$ are fixed to the variations
$\phi_a^+=\frac{\delta\Psi}{\delta \phi^a}$. This can be
reformulated in terms of the BV symplectic structure on the
space of superfields $\mbf\cM$. For this, we consider the case where
the source dg-manifold is the superworldvolume $\cW=T[1]\Sigma_d$ with
local coordinates $\hat z=(\sigma,\theta)$ and write a generic BV symplectic structure on superfields in its canonical form as
\be \label{eq:AKSZgenBVsymplGF}
\bs{\omega} \, = \int_{{T}[1]\Sigma_d} \,\dd^d\hat z\ \, \bs{\delta}
\bs{\phi}^+_a(\hat z)\, \bs{\delta} \bs{\phi}^a(\hat z) \ , 
\ee
where we chose a convenient ordering of antifields $\bs{\phi}_a^+$ and
fields $\bs{\phi}^a$ in this way. We write $|a|$ for the degree of the
superfield $\bs{\phi}^a$; then its antifield $\mbf\phi_a^+$ has degree
$d-1-|a|$. If the Liouville potential is chosen as 
\be
\bs{\vartheta} \, = \, \int_{T[1]\Sigma_d}\, \dd^d \hat{z} \
\bs{\phi}^+_a(\hat z)\,  \bs{\delta} \bs{\phi}^a(\hat z) \ ,
\ee
then the kinetic part of the AKSZ action is 
\be
\bs{S}_0 \, = \, \int_{T[1]\Sigma_d}\, \dd^d \hat{z} \ (-1)^{|a|} \,
\bs{\phi}^+_a(\hat z)\, \bs{D} \bs{\phi}^a(\hat z) \ .
\ee

We choose a gauge fixing fermion $\bs{\Psi}[\bs{\phi}]$, which is a
functional on superfields $\bs{\phi}(\hat z)\in\mbf\cM$, and fix the antifields to
\be \label{eq:AKSZgenGFAntifields}
\bs{\phi}^+_a (\hat z) \, = \, (-1)^{|a|\,(d+1)}\, \frac{\rdelta
  \bs{\Psi}}{\delta \bs{\phi}^a(\hat z)} \ ,
\ee
where an extra sign factor has been introduced, which depends on the dimension of the worldvolume. The left-acting functional derivative is defined in the usual way by
\be
\lim_{\epsilon \rightarrow 0}\, \frac{\bs{\Psi}[\bs{\phi} + \epsilon\,
  \mbf\xi\,] \, - \, \bs{\Psi}[\bs{\phi}]}{\epsilon} \, =: \,
\int_{T[1]\Sigma_d} \,\dd^d\hat z \ \mbf\xi(\hat z) \,
\frac{\rdelta \bs{\Psi}}{\delta \bs{\phi}(\hat z)} \ . 
\ee
The BV symplectic form \eqref{eq:AKSZgenBVsymplGF} in the gauge that
is fixed
by $\bs{\Psi}[\bs{\phi}]$ according to \eqref{eq:AKSZgenGFAntifields} is
\be \begin{aligned}
\bs{\omega}_{\bs{\Psi}} \, & = \sum_a\, (-1)^{|a|\,(d+1)} \
\int_{T[1]\Sigma_d} \,\dd^d\hat z_1 \ \bs{\delta} \frac{\rdelta
  \bs{\Psi}}{\delta \bs{\phi}^a(\hat z_1)}\, \bs{\delta} \bs{\phi}
^a(\hat z_1) \\[4pt]
& = \, (-1)^{d+1}\, \sum_{a,b}\, (-1)^{|b|\,(|a|+1)+|a|\,d} \,
\int_{T[1]\Sigma_d} \,\dd^d\hat z_1 \, \int_{T[1]\Sigma_d} \,\dd^d\hat
z_2 \ \frac{\rdelta{}^2 \bs{\Psi}}{\delta \bs{\phi}^b(\hat z_2)\,
  \delta \bs{\phi}^a(\hat z_1)}\, \bs{\delta} \bs{\phi}^b(\hat z_2) \,
\bs{\delta} \bs{\phi}^a(\hat z_1) \ .
\end{aligned}\ee
Interchanging variables and indices yields sign changes which are
given by
\be \begin{aligned}
\frac{\rdelta{}^2 \bs{\Psi}}{\delta \bs{\phi}^b(\hat z_2)\,\delta
  \bs{\phi}^a(\hat z_1)} \, &= \, (-1)^{(|a|+d)\,(|b|+d)} \,
\frac{\rdelta{}^2 \bs{\Psi}}{\delta \bs{\phi}^a(\hat z_1)\,\delta
  \bs{\phi}^b(\hat z_2)} \ , \\[4pt]
\bs{\delta} \bs{\phi}^b(\hat z_2) \, \bs{\delta} \bs{\phi}^a(\hat z_1)
\, &= \, (-1)^{(|a|+1)\,(|b|+1)} \, \bs{\delta} \bs{\phi}^a(\hat z_1)
\, \bs{\delta} \bs{\phi}^b(\hat z_2) \ , \\[4pt]
\int_{T[1]\Sigma_d} \,\dd^d\hat z_1 \ \int_{T[1]\Sigma_d} \,\dd^d\hat
z_2 \,  &= \, (-1)^{d} \, \int_{T[1]\Sigma_d} \,\dd^d\hat z_2 \ \int_{T[1]\Sigma_d} \,\dd^d\hat z_1 \ .
\end{aligned} \ee
This shows that the gauge fixed BV symplectic form is a product of a
symmetric and an antisymmetric expression, and hence
$\bs{\omega}_{\bs{\Psi}}=0$. Thus gauge fixing with a fermion in the
sense of \eqref{eq:AKSZgenGFAntifields} means that one chooses a
Lagrangian submanifold $\mbf\cL$ of the space of all fields $\mbf\cM$,
i.e.~ a subspace $\mbf\cL\subset\mbf\cM$ on which the symplectic form
$\mbf\omega$ vanishes and which has half the dimension of
$\mbf\cM$. In the following we use this prescription generally: A choice
of gauge in BV quantization is equivalent to a choice of a Lagrangian
submanifold $\mbf\cL$ in $\mbf\cM$. The Batalin-Vilkovisky theorem~\cite{Batalin1981}
ensures that the path integral over $\mbf\cL$ is independent of the
choice of representative for the homology class of the Lagrangian submanifold $\mbf\cL$. By the localization theorem, the path integral localizes over the fixed point locus of the BV--BRST charge $\mbf Q$ in the Lagrangian subspace $\mbf\cL$.
From a physical point of view, the Lagrangian submanifold intersects
the gauge orbits orthogonally, i.e. the action of the
BV--BRST charge $\bv{\bs{S}}{\, \cdot \,}$ vanishes on Lagrangian submanifolds,
as the BV bracket acts as zero there. Thus the BV gauge symmetry is
completely fixed on Lagrangian submanifolds.\footnote{We have not
  studied the Gribov problem in this context.}

Let us now reformulate these observations in terms of the expansion
coefficients of superfields. An arbitrary superfield $\mbf\phi^a$ can be expanded in terms of the degree~1 fiber coordinates $\theta^\mu$ of $\cW=T[1]\Sigma_d$ in the form
\be \label{eq:superfieldexp}
\bs{\phi}^a(\hat z) \, = \, \phi^{(0) \, a} (\sigma) \, + \, \phi^{(1) \, a}_{\mu_1}
(\sigma) \, \theta^{\mu_1} \, + \, \frac 12\, \phi^{(2) \, a}_{\mu_1 \mu_2}
(\sigma) \, \theta^{\mu_1} \, \theta^{\mu_2} \, + \, \cdots \, + \,
\frac{1}{d!}\, \phi^{(d) \, a}_{\mu_1 \cdots \mu_d} (\sigma) \, \theta^{\mu_1} \cdots \theta^{\mu_d} \ ,
\ee
where $\phi^{(p) \, a}$ are the degree $|a|-p$ coefficients of $\bs{\phi}^a$ which can be identified with $p$-forms on $\Sigma_d$.
The BV symplectic form can be written as an integral over the original
worldvolume $\Sigma_d$ as
\be \begin{aligned}
\bs{\omega} \, &= \, \sum_{p=0}^d  \ \int_{\Sigma_d}\,  \bs{\delta} \phi^{(p) \, +}_{a} \w \bs{\delta} \phi^{(p) \, a} \\[4pt]
 \, & = \, \sum_{p=0}^d \, \frac{1}{p!} \ \int_{\Sigma_d}\, \dd^d
 \sigma \ \sum_a\, (-1)^{|a|+p} \,  \bs{\delta} \wt{\phi}{}^{\,(p) \, + ; \, \mu_1 \cdots \mu_p}_{a} (\sigma) \, \bs{\delta} \phi^{(p) \, a}_{\mu_{1}\cdots \mu_{p}} (\sigma) \ ,
\end{aligned} \ee 
where $\wt{\phi}{}^{\,(p) \, +}_{a}$ is the dual antifield of
$\phi^{(p) \, a}$ defined by
\be
\wt{\phi}{}^{\,(p) \, + ; \, \mu_{d-p+1} \cdots \mu_d}_{a} \, = \, (-1)^{d\,(d+1+|a|+p)+|a|\,(p+1)+p} \, \frac{1}{(d-p)!} \, \epsilon^{\mu_1 \cdots \mu_d} \, \big({\phi^{+}_{a}}\big)^{(d-p)}_{\mu_1 \cdots \mu_{d-p}}  \ .
\ee
Here $\epsilon^{\mu_1 \dots \mu_d}$ is the Levi-Civita tensor density
on $\Sigma_d$, and $\big({\phi^{+}_{a}}\big)^{(d-p)}$ are the
expansion coefficients of the superfield $\bs{\phi}^+_a$. The BV symplectic form with this sign convention gives the canonical Poisson bracket relations
\be
\big\{ \phi^{(p) \, a}_{\mu_1 \cdots \mu_p} \,,\, \wt{\phi}{}^{\,(p'\,) \, + ; \, \nu_1 \cdots \nu_{p'}}_{b} \big\} \, = \, \delta^{pp'} \ \delta^a{}_b \ \delta_{\mu_1 \cdots \mu_p}{}^{\nu_1 \dots \nu_{p'}} \ .
\ee
Gauge fixing with a fermion $\Psi[\phi]$ then takes the more familiar form
\be
\wt{\phi}{}^{\,(p) \, +}_a \, = \, \frac{\rdelta \Psi}{\delta
  \phi^{(p) \, a}} \ ,
\ee
and it gives a vanishing symplectic structure
$\bs{\omega}_{\bs{\Psi}}=0$, whence the fermion $\Psi$ generates a
Lagrangian submanifold in the terms of the expansion coefficients as well.

\subsection{$B$-fields and the Poisson sigma-model}
\label{sec:AKSZPoisson}

In dimension $d=2$, the AKSZ theory with target space a degree~1
QP-manifold describes the topological sigma-model for closed strings
in an NS--NS $B$-field background. In
the worldsheet sigma-model approach, the fundamental field is a map
$X: \Sigma_2 \rightarrow M$ from a closed and oriented Riemann
surface $\Sigma_2$ to a target space $M$. Denoting the local
coordinates by $( X^i)\in M$ and $(\sigma^\mu)\in\Sigma_2$, the string
field $X$ is described by a set of functions $\big(X^i(\sigma^\mu)\big) $ on $\Sigma_2$. The topological part of the bosonic string action is 
\be
I_{\Sigma_2,\, B} \, = \, \int_{\Sigma_2}\, X^*(B) \, = \, \frac 12\, \int_{\Sigma_2}\, B_{ij} \, \dd X^i \wedge \dd X^j \ ,
\ee
where $B=\frac 12\, B_{ij} \, \dd X^i \wedge \dd X^j$ is the
Kalb-Ramond two-form field on $M$. If $B$ is non-degenerate, it corresponds to an almost symplectic structure on $M$ and we can write the classically equivalent first order string sigma-model
 \be\label{eq:Poissonaction}
 I^{(1)}_{\Sigma_2,\, \pi} \, = \, \int_{\Sigma_2}\, \Big( \chi_i \wedge \dd X^i \, + \, \frac 12 \, \pi^{ij} \,  \chi_i \wedge \chi_j \Big) \ ,
 \ee
where $B_{ij}$ is the inverse of $-\pi^{ij}$ and
$\chi=(\chi_i)\in\Omega^1(\Sigma_2,X^*T^*M)$ is an auxiliary
one-form. The bivector $\pi=\frac 12\, \pi^{ij} \,
\frac{\partial}{\partial X^i} \wedge \frac\partial{\partial X^j}$ on
$M$ is a Poisson bivector on-shell, which is equivalent to a flat
$B$-field $\dd B = 0$, so that the Kalb-Ramond field corresponds to a
symplectic structure on $M$. This is the action functional of the
Poisson sigma-model~\cite{Ikeda1993,Schaller1994}. 

The AKSZ formulation of the Poisson sigma-model is studied in \cite{Cattaneo2001}.
We take $\cW= T [1]\Sigma_2$, and $\cM= T ^*[1]M$ with\footnote{In the
  present context we use the terminology `degree' to mean ghost
  number. In general the degree means total degree, which is the sum
  of the ghost number and the form degree on the dg-manifold, but here we have only functions.} degree~0 base coordinates $X^i$ on $M$ and degree~1 fiber coordinates $\chi_i$. The canonical symplectic form on $\cM$ is 
\be
\omega=\dd \chi_i \wedge \dd X^i \  ,
\ee
which leads to the canonical graded Poisson bracket $\{ \chi_i , X^j \} = \delta_i{}^j$ on the local coordinates of $\cM$.
We choose the Liouville potential to be $\vartheta= \chi_i\, \dd
X^i$. The most general form of a degree~2 Hamiltonian function
$\gamma$ on
$\cM$ is given by a (possibly degenerate) $(0,2)$-tensor $\pi=\pi^{ij}\, \frac\partial{\partial X^i}\otimes\frac\partial{\partial X^j}$ on $M$ as
\be
\gamma \, = \, \frac 12 \, \pi^{ij}(X) \, \chi_i \, \chi_j \ .
\ee 
The corresponding cohomological vector field $Q_\gamma$ on $\cM$ is 
\be\label{eq:QgammaPoisson}
Q_\gamma\, = \, \pi^{ij}\,\chi_j \, \frac{\partial}{\partial X^i} \, + \, \frac 12 \, \frac{\partial\pi^{ij}}{\partial X^k}\, \chi_i \, \chi_j \, \frac{\rd}{\partial\chi_k} \ .
\ee
Compatibility of $Q_\gamma$ with $\omega$ implies $\pi\in\Gamma(\bigwedge^2TM)$ and 
the classical master equation $\{ \gamma, \gamma \}=0$ implies that $\pi= \frac 12 \,\pi^{ij} \, \frac{\partial}{\partial X^i}\wedge \frac\partial{\partial X^j}$ must be a Poisson bivector on $M$, i.e. $\pi^{l[i}\,\frac{\partial\pi^{jk ]}} {\partial X^l} = 0$. In other words, a symplectic Lie 1-algebroid is the same thing as a Poisson manifold $(M,\pi)$, which by construction is also a Lie algebroid on the cotangent bundle $T^*M$. 
The Hamiltonian function determines a derived bracket which defines a Poisson bracket on $C^\infty(M)$ through
\be
\{ f,g \}_{\pi} \, = \, \pi^{ij} \, \frac{\partial f}{\partial X^i} \, \frac{\partial g}{\partial X^j} \, = \, - \, \{ \{ f , \gamma \} , g \} \  .
\ee

The kinetic part $\bs{S}_0$ of the AKSZ action is inherited from the cohomological vector field $Q_{\cW}$ on $ \cW=T [1]\Sigma_2$, and is given by
\be
\bs{S}_0 \, = \, \int_{ T [1]\Sigma_2} \,\dd^2\hat z \ \bs{\chi}_i \, \bs{D}  \bs{X}^i \ ,
\ee
where as before the superworldsheet differential is $\bs{D}=\theta^\mu\, \frac\partial{\partial\sigma^\mu}=Q_\cW$. The BV bracket has the form
\be
\bv{\,\cdot\,}{\,\cdot\,} \, = \, \int_{ T [1]\Sigma_2}\,\dd^2\hat z \
\frac\delta{\delta\bs{X}^i} \, \w \, \frac\delta{\delta\bs{\chi}_i} \ ,
\ee
where
\be
\frac\delta{\delta\bs{X}^i} \, \w \, \frac\delta{\delta\bs{\chi}_i} :=
\frac\ldelta{\delta\bs{X}^i} \, \frac\rdelta{\delta\bs{\chi}_i} \, -
\, \frac\ldelta{\delta\bs{\chi}_i} \, \frac\rdelta{\delta\bs{X}^i}  \ .
\ee
Together these ingredients give the AKSZ action for the Poisson sigma-model as
\be \label{eq:PoissonActAKSZ}
\bs{S} \, = \, \int_{ T [1]\Sigma_2}\,\dd^2\hat z \ \Big(\, \bs{\chi}_i \, \bs{D}  \bs{X}^i \, + \, \frac 12 \, \bs{\pi}^{ij} \, \bs{\chi}_i\, \bs{\chi}_j \Big) \ ,
\ee
where $\mbf f=\mbf\phi^*(f)=f(\mbf\phi)$ for a function $f$ on $\cM$
and $\mbf\phi\in\mbf\cM$.
Integrating over the odd coordinates $\theta^\mu$ and restricting to
the degree~0 fields in \eqref{eq:PoissonActAKSZ} recovers the
classical action \eqref{eq:Poissonaction}, and in this sense the action
\eqref{eq:PoissonActAKSZ} provides a BV quantization of the original
Poisson sigma-model. By the fixed point theorem, the path integral
localizes onto critical points of the action $\mbf S$, which are the
fixed points of the cohomological vector field $Q_\gamma$ that
defines the Poisson cohomology $H^\bullet_\pi(M)$ of $M$. This
formulation of the Poisson sigma-model also gives an AKSZ construction of
the A-model~\cite{AKSZ1997,Park2000,Ikeda2012,Pestun2006,Zucchini2004,Stojevic2005}.

\subsection{Courant algebroids and topological membranes}
\label{sec:CourantAlgAKSZ}

The next dimension $d=3$ is particularly relevant to extending the
Poisson sigma-model to closed string backgrounds with non-zero NS--NS three-form
flux $H=\dd B$, or to M-theory backgrounds with three-form $C$-field. In this setting the closed strings are replaced with
membranes described by maps $X=(X^i)$ from a closed three-dimensional
worldvolume $\Sigma_3$ to the target space $M$. The topological part
of the bosonic membrane action is the Wess-Zumino coupling
\be\label{eq:3H}
I_{\Sigma_3,H} \, = \, \int_{\Sigma_3}\, X^*(H) \, = \, \frac{1}{3!}\,\int_{\Sigma_3}\, H_{ijk} \, \dd X^i \w \dd X^j \w \dd X^k \ .
\ee
This action is classically equivalent to the first order membrane sigma-model action
\be \label{eq:Membr1}
I_{\Sigma_3,H}^{(1)} \, = \, \int_{\Sigma_3}\,\Big( F_i \w \big(\dd X^i - \psi^i \big) \, - \, \chi_i \w \dd \psi^i \, + \, \frac{1}{3!} \, H_{ijk} \, \psi^i \w \psi^j \w \psi^k \Big) \ ,
\ee
where $\psi=(\psi^i)\in \Omega^1(\Sigma_3,X^*TM)$ and $\chi=(\chi_i)\in\Omega^1(\Sigma_3,X^*T^*M)$ are one-forms, while $F=(F_i)\in\Omega^2(\Sigma_3,X^*T^*M)$ is an auxiliary two-form. 
The corresponding AKSZ sigma-model is defined on worldvolume superfields
with target space a QP-manifold of degree~2, which corresponds to a
Courant algebroid~\cite{Roytenberg2002b}. 

Recall that a Courant
algebroid on a manifold $M$ is a vector bundle $E$ over $M$ equiped
with a symmetric non-degenerate bilinear form $\langle \, \cdot \, ,
\, \cdot \, \rangle$ on its fibers, an anchor map $\rho: E \rightarrow
T M$, and a binary bracket of sections $[\, \cdot \, , \, \cdot \,
]_{\rm D}$, called the Dorfman bracket, which together satisfy
\be\label{eq:CourantAx}
\begin{aligned}
{}[ e_1 ,[e_2,e_3 ]_{\rm D} ]_{\rm D} \, &= \ [[e_1,e_2]_{\rm D},e_3]_{\rm D} \, + \, [e_2,[e_1,e_3]_{\rm D}]_{\rm D} \  , \\[4pt]
\rho(e_1) \langle e_2,e_3 \rangle \, &= \ \langle [e_1,e_2]_{\rm D} , e_3 \rangle \, + \, \langle e_2 , [e_1,e_3]_{\rm D} \rangle \ , \\[4pt]
\rho(e_1) \langle e_2,e_3 \rangle \, &= \ \langle e_1 , [e_2, e_3 ]_{\rm D} \, + \, [e_3 , e_2]_{\rm D} \rangle \ , 
\end{aligned}
\ee
where $e_1,e_2,e_3$ are sections of $E$.
The most common example is the standard Courant algebroid, which
features in generalized geometry~\cite{Hitchin2004,Gualtieri2003}. It is an extension of the Lie
algebroid of tangent vectors by cotangent vectors and is based on
the generalized tangent bundle
\be
E= T M\oplus T ^*M \ ,
\ee
with the three operations
   \be \begin{aligned}
 \langle A + \alpha , B +  \beta \rangle \, &= \ \iota_A\beta \, + \, \iota_B \alpha \ , \\[4pt]
 \rho(A + \alpha) \, &= \ A \ , \\[4pt]
 [A+\alpha,B+\beta]_{\rm D}\, &= \ [A,B] \, + \, \cL_A \beta \, - \, \iota_B\, \dd \alpha \ ,  \label{eq:DorfmanBr}
 \end{aligned} \ee
where the sections of $E= T M\oplus T ^*M$ are composed of vector
fields $A,B$ and one-forms $\alpha,\beta$.
The NS--NS $H$-flux then appears as a twisting of the standard Courant
algebroid, which gives rise to a deformation of the Dorfman bracket
through an extra term as
 \be \label{eq:CourantBrTwistedH}
 [A+\alpha,B+\beta]_{{\rm D},H}\, = \ [A,B] \, + \, \cL_A \beta \, -
 \, \iota_B\, \dd \alpha \, + \, \iota_A \iota_B H \ .
\ee

Let us now review the correspondence between Courant algebroids and
the AKSZ construction, following~\cite{Ikeda2012,Heller2016} for the most part. Given a QP-manifold $\cM$ of degree~2, we choose local Darboux
coordinates $(X^i, \zeta^a, F_i)$ with degrees $(0,1,2)$ in which the
graded symplectic structure is given as 
\be
\omega \, = \, \dd F_i \w \dd X^i \, + \, \frac{1}{2}\, k_{ab} \, \dd \zeta^a \w \dd \zeta^b \ .
\ee
Here we have introduced a constant metric $k_{ab}$ on the degree~1 subspace,
which is a local coordinate expression of the symmetric pairing in the
corresponding Courant algebroid. The graded Poisson brackets of the
coordinates are canonical in the sense that
\be
\{ X^i , F_j \} \, = \, \delta^i{}_j \qquad \mbox{and} \qquad \{ \zeta^a , \zeta^b \} \, = \, k^{ab} \ ,
\ee
where $k^{ab}$ is the inverse of $k_{ab}$. The most general form of
the degree~3 Hamiltonian function $\gamma$ is given by
\be
\gamma \, = \, \rho^i{}_a (X) \, F_i \, \zeta^a \, + \, \frac{1}{3!} \,
T_{abc}(X) \, \zeta^a \, \zeta^b \, \zeta^c \ ,
\ee
where the functions $\rho^i{}_a $ and $T_{abc}$ on $M$ give the
local forms of the anchor map and fluxes, respectively. 
The three operations on the Courant algebroid are given by taking
derived brackets defined by $\gamma$ 
and the graded Poisson bracket through
 \be 
[e_1,e_2]_{\rm D} =  \{\{ e_1 , \gamma \} , e_2 \} \ , \qquad
\langle e_1, e_2\rangle = \{ e_1,e_2 \} \qquad \mbox{and} \qquad
\rho(e) =  \{ e , \{ \gamma , \, \cdot \, \}\} \ .
\ee
These operations are defined on degree~1 functions $e$ with local
expression $e=f_a(X)\,\zeta^a$, where $f_a$ is a degree~0
function on the body $M=\cM_0$ of $\cM$, which are identified as local
sections of a vector bundle $E$ over $M$. They satisfy the Courant algebroid axioms in \eqref{eq:CourantAx} as a
consequence of the classical master equation $\{\gamma,\gamma
\}=0$. 

Conversely, given a Courant
algebroid on a vector bundle $E$ over $M$, we define the target QP-manifold $\cM$ of
degree~2 to be the symplectic 
submanifold of $T^*[2]E[1]$ corresponding to the isometric embedding
$E\hookrightarrow E\oplus E^*$ with respect to the Courant algebroid
pairing and the canonical dual pairing. Then $X^i$ are local
coordinates on $M$, $F_i$ are local fiber coordinates of the shifted
cotangent bundle $T^*[2]M$, and $\zeta^a$ are local fiber coordinates
of the shifted vector bundle $E[1]$. In other words, a symplectic
Lie 2-algebroid is the same thing as a Courant algebroid. 

In this paper we shall only deal with the standard Courant algebroid on the
generalized tangent bundle $E= T M \oplus  T ^*M$. The corresponding
QP-manifold of degree~2 is then simply $\cM= T ^*[2] T [1]M$. The local
degree~1 coordinates are dual pairs $\zeta^a=(\psi^i,\chi_i)$ and the symplectic form is
\be \label{eq:StdCourantSympl} 
\omega \, = \, \dd F_i \w \dd X^i \, + \, \dd \chi_i \w \dd \psi^i \ .
\ee
For the Liouville potential we choose $\vartheta=F_i\, \dd
X^i-\chi_i\, \dd\psi^i$. The simplest Hamiltonian function with $\rho^i{}_a = \delta^i{}_a$ and
$T_{abc}=0$ is given by
\be \label{eq:HamStdCourant}
\gamma_0 \, = \, F_i \, \psi^i \ , 
\ee
and the corresponding cohomological vector field 
\be 
Q_{\gamma_0} = \psi^i\, \frac{\partial}{\partial X^i}
\ee
corresponds to the de~Rham differential on $M$. Its derived brackets on degree 1 functions
\be
 A^i \,\chi_i \, + \, \alpha_i \, \psi^i \ \longleftrightarrow \ A^i
 \,\frac\partial{\partial X^i} \, + \, \alpha_i \, \dd X^i 
\ee
yields the standard Courant bracket which is the antisymmetrization of
the standard Dorfman bracket in \eqref{eq:DorfmanBr} given by
\be \label{eq:CourantBr}
[ A + \alpha , B + \beta ]_{\rm C} \, = \, [A,B]  +  \cL_A \beta -  \cL_B \alpha  -  \frac 12 \, \dd (\iota_A \beta  - \iota_B \alpha) \ .
\ee
The Courant bracket is the natural bracket in generalized geometry
which is compatible with the commutator
algebra 
of generalized Lie derivatives~\cite{Hitchin2004,Gualtieri2003}.
Only
the simplest case of pure NS--NS flux $T_{abc}=H_{ijk}$ is consistent with the choice of
anchor map $\rho^i{}_a = \delta^i{}_a$ of the standard Courant
algebroid, which is necessarily closed by the classical master equation. Given a Kalb-Ramond two-form field $B$ on $M$, with $H=\dd
B$, canonical transformation of the Hamiltonian function
\eqref{eq:HamStdCourant} by the degree~2 function $B=\frac12\,
B_{ij}(X)\, \psi^i\,\psi^j$ on $\cM$ yields the twisted Hamiltonian function
\be
\gamma_H \, := \, \e^{\delta_B}\gamma_0 \, = \, F_i \, \psi^i \, + \, \frac{1}{3!} \, H_{ijk} \,
\psi^i \, \psi^j \, \psi^k \ ,
\ee
which gives the twisted Courant bracket as the antisymmetrization of
 \eqref{eq:CourantBrTwistedH}.

It is evident from the general construction that Courant algebroids
are uniquely encoded (up to isomorphism) in the corresponding AKSZ topological membrane theories, which are called Courant sigma-models~\cite{Roytenberg2007}. In the particular example of the standard Courant algebroid on $E=TM\oplus T^*M$ twisted by a closed NS--NS three-form flux $H$, the mapping space $\mbf\cM$ of superfields supports the canonical BV symplectic structure 
\be \label{eq:StdCourantBVomega}
\mbf\omega = \int_{T[1]\Sigma_3}\,\dd^3\hat z \ \big(\mbf\delta\mbf
X^i\, \mbf\delta\mbf F_i + \mbf\delta\mbf\psi^i\, \mbf\delta\mbf\chi_i\big) \ ,
\ee
where the ghost number $U=2$ superfields $\mbf F_i$ and $U=0$ superfields $\mbf X^i$, as well as the conjugate pairs of
$U=1$ superfields $\mbf\chi_i$ and $\mbf\psi^i$, contain each other's antifields respectively. The AKSZ construction leads to the action
\be \label{eq:AKSZactionStdCourant}
\mbf S = \int_{T[1]\Sigma_3}\,\dd^3\hat z \ \Big(\mbf F_i\, \mbf D\mbf X^i -
\mbf\chi_i\, \mbf D\mbf\psi^i + \mbf F_i\, \mbf\psi^i + \frac1{3!}\, \mbf H_{ijk}\, \mbf\psi^i\,\mbf\psi^j\,\mbf\psi^k\Big) \ ,
\ee
which solves the classical master equation $(\mbf S,\mbf
S)_{\mathrm{BV}}=0$. Integrating over $\theta^\mu$ and restricting to degree~0 fields in \eqref{eq:AKSZactionStdCourant} recovers the classical action \eqref{eq:Membr1}.

\subsection{Lie algebroids up to homotopy and topological threebranes}
\label{sec:LieAlguth}

Just as it proves useful to view closed strings as modes of
membranes when deforming their target spaces by fluxes, it is
likewise useful to view membranes as modes of higher
degrees of freedom, threebranes, particularly when the membranes are regarded as the
fundamental objects in M-theory with background four-form fluxes
$G=\dd C$. With notation as previously, the threebrane theory is defined on a four-dimensional worldvolume $\Sigma_4$, and the topological part of the bosonic action is
\be \label{eq:TopHFluxTerm}
I_{\Sigma_4,G} \, =\,\int_{\Sigma_4}\, X^*(G)\,=\, \frac{1}{4!}\, \int_{\Sigma_4}\, G_{ijkl}\,\dd X^i \w\dd X^j \w\dd X^k \w\dd X^l \ .
\ee
This action is classically equivalent to the first order threebrane sigma-model action
\be\label{eq:Phys4Brane}
I^{(1)}_{\Sigma_4,G} \, = \, \int_{\Sigma_4}\,  \Big(F_i \w \big(\psi^i - \dd X^i\big) + \chi_i \w \dd \psi^i + \frac{1}{4!}\, G_{ijkl}\, \psi^i \w \psi^j \w \psi^k \w \psi^l\Big) \  ,
\ee
where $\psi\in\Omega^1(\Sigma_4,X^*TM)$ and $\chi\in\Omega^2(\Sigma_4,X^* T^*M )$, while $F\in\Omega^3(\Sigma_4,X^* T^*M )$ is an auxiliary three-form. 
In dimension $d=4$, the target superspace of the AKSZ
construction is a QP-manifold of degree 3, which is equivalent to a
higher algebroid structure introduced in \cite{Ikeda2011} that arises
from a homotopy deformation of a Lie algebroid. It is called a Lie algebroid up to homotopy. 

Let $E_0$ be a vector bundle over a manifold $M$. We consider a
general QP-manifold of degree~3 on $\cM= T ^*[3]E_0[1]$, regarded as a symplectic Lie 3-algebroid on $E_0$ with underlying N-manifold
\be 
\cM \ = \ M\longleftarrow E_0[1] \longleftarrow E_0[1]\oplus E_0^*[2]\longleftarrow T^*[3]E_0[1] \ .
\ee
The local
coordinates on $\cM$ are denoted $(X^i,\psi^a, \chi_a, F_i)$ with
degrees $(0,1,2,3)$, where $X^i$ are local coordinates on $M$,
$\psi^a$ are local fiber coordinates of the shifted vector bundle $E_0[1]$, $\chi_a$ are dual
fiber coordinates of $T^*[3]E_0[1]\to E_0[1]$, and $F_i$ are local
fiber coordinates of the shifted cotangent bundle $T^*[3]M$. The
canonical symplectic structure is given by
\be
\omega \, = \, \dd F_i \w \dd X^i \, + \, \dd\psi^a \w \dd \chi_a \ .
\ee
The most general form of a degree~4 Hamiltonian function $\gamma$  on
$\cM$ is
given by a sum 
\be \label{eq:LieAlguthHamDecomp}
\gamma \, = \, \gamma_k \, + \, \gamma_{\rho} \, + \, \gamma_T \  ,
\ee
where
\be \begin{aligned} \label{eq:LieAlguthHamParts}
\gamma_k \, =& \ \frac 12\, {k}^{ab}(X) \, \chi_a \, \chi_b \  , \\[4pt]
\gamma_{\rho} \, = & \ {{\rho}^i}_a (X)\,  F_i \, \psi^a \, + \, \frac 12\, {{f}^a}_{bc}(X) \, \chi_a \, \psi^b \, \psi^b \  , \\[4pt]
\gamma_T \, = & \ \frac{1}{4!}\, {T}_{abcd}(X) \, \psi^a \, \psi^b \,
\psi^c \, \psi^d \ ,
\end{aligned}\ee
are given by functions ${k}^{ab}$, ${{\rho}^i}_a$, ${{f}^a}_{bc}$ and
${T}_{abcd}$ on $M$. A Lie algebroid up to homotopy is defined with
respect to this decomposition of the Hamiltonian function as the vector bundle $E_0$ over $M$ with a symmetric pairing $\langle\,\cdot\,,\,\cdot\,\rangle$ on sections of $E_0^*$, an anchor map $\rho:\, E_0 \, \rightarrow \, TM$, an antisymmetric bracket $[\, \cdot \, , \, \cdot \,]_{\rm{uth}}$ on sections of $E_0$, a de~Rham-type differential $\mathsf{d}$ on sections of $\bigwedge^\bullet E_0$, and a four-form $\Omega$ on $E_0$. We can identify sections $e$ of $E_0$ with degree~2 functions $e=f^a (X) \, \chi_a$ and sections $\alpha$ of $E_0^*$ with degree~1 functions $\alpha=g_a (X) \, \psi^a$, where $f^a$ and $g_a$ are degree~0 functions on $M$. Then the five operations are defined via derived brackets as
\be 
\begin{aligned}
\langle\alpha_1,\alpha_2\rangle \, &= \, \{ \{ \gamma_k , \alpha_1 \} , \alpha_2 \} \ , \\[4pt]
\rho(e) \, &= \, \{ \{ \gamma_{\rho}, e\} , \, \cdot \, \} \ , \\[4pt]
[e_1,e_2]_{\rm{uth}} \, &= \, \{ \{ \gamma_{\rho}, e_1\} , e_2 \} \ , \\[4pt]
\mathsf{d} \, &= \, \{\gamma_{\rho}, \, \cdot \, \} \ , \\[4pt]
\Omega(e_1,e_2,e_3,e_4) \, &=  \, \{ \{ \{ \{ \gamma_T , e_1\} , e_2 \} , e_3 \} , e_4 \} \ .
\end{aligned}
\ee
The pairing additionally defines a symmetric bundle map $\mathfrak{d} : E^* \rightarrow E$ by
\be
\mathfrak{d} \alpha	 \, = \, - \, \{ \gamma_k , \alpha \} \ .
\ee

The classical master equation $\{\gamma,\gamma\}=0$ implies that these operations obey the identities
\be \label{eq:uthidentities}
\begin{aligned}
{}[e_1,f \, e_2]_{\rm{uth}} \, &= \, f\, [e_1,e_2]_{\rm{uth}} \, + \, \big(\rho(e_1)f\big) \, e_2 \qquad \mbox{for} \quad f\in C^\infty(M) \ , \\[4pt]
[ [e_1, e_2]_{\rm{uth}}, e_3 ]_{\rm{uth}} \, + \, \rm{cyclic} \, &= \, \mathfrak{d}\,\Omega(e_1,e_2,e_3,\,\cdot\,) \ , \\[4pt]
\rho\circ\mathfrak{d} \, &= \, 0 \ , \\[4pt]
\rho(e) \langle\alpha_1,\alpha_2\rangle \, &= \, \langle \cL_e \alpha_1, \alpha_2\rangle \, + \, \langle\alpha_1, \cL_e \alpha_2\rangle \qquad \mbox{with} \quad \cL_e\,:=\,\{\{ \gamma_{\rho},e\},\, \cdot \, \} \ , \\[4pt]
\mathsf{d}\circ\Omega \, &= \, 0 \ ,
\end{aligned}
\ee
and we also note that $\mathsf{d}^2\neq 0 $ in general. In other words, a symplectic Lie 3-algebroid is a vector bundle with operations $([\,\cdot\, ,\, \cdot\,]_{\rm{uth}}, \rho,\mathfrak{d},\Omega)$ characterized by the algebraic identities \eqref{eq:uthidentities}. A particularly interesting feature behind the algebraic structure of a Lie algebroid up to homotopy is that its bracket can be extended to all degree 2 functions on $ \cM=T ^*[3]E_0[1]$, which are identified as the sections of $E:=E_0\oplus\bigwedge^{2}E_0^*$. This leads to a higher analogue of the Courant bracket
\be
[\, \cdot \, , \, \cdot \,]_{\rm{2C}} \, = \, \{\{\gamma ,\, \cdot \, \}, \, \cdot \, \} \  , 
\ee
where now the full Hamiltonian function is used. We shall call it a
2-Courant bracket in the following.

The simplest relevant example for us is what we shall call the \emph{standard} Lie algebroid up to homotopy, which is the case $E_0= T M$. The symplectic structure is
\be
\omega \, = \, \dd X^i \w \dd  F_i \, + \, \dd  \psi^i \w \dd  \chi_i \ .
\ee
We choose the Liouville potential given by $\vartheta=F_i\,\dd X^i + \chi_i\, \dd \psi^i$. The simplest Hamiltonian function from \eqref{eq:LieAlguthHamDecomp} and \eqref{eq:LieAlguthHamParts} has identity anchor map $\rho^i{}_j=\delta^i{}_j$ with all other structure functions equal to zero, and is given by
\be
\gamma_0 \, = \,  F_i \, \psi^i \  .
\ee
The cohomological vector field is again the de~Rham vector field $Q_{\gamma_0} = \psi^i\, \frac{\partial}{\partial X^i}$ on $\cM=T^*[3]T[1]M$. 
In this instance, the derived bracket on degree~2 functions
\be 
A^i\, \chi_i + \frac12\,\lambda_{ij}\, \psi^i\, \psi^j \ \longleftrightarrow \ A^i\, \frac\partial{\partial X^i} + \frac12\, \lambda_{ij}\, \dd X^i\wedge\dd X^j
\ee
gives the standard 2-Courant bracket on 
the vector bundle
\be 
E=T M\oplus \mbox{$\bigwedge^{2}$}\, T ^*M \ ,
\ee
which reads explicitly as
\be \label{eq:StdHigherCourantBr}
[A + \lambda, B + \xi]_{\rm{2C}} \, = \, [A,B] \, + \, \cL_A \xi \, - \, \cL_B \lambda \, + \, \frac 12 \, \dd (\iota_B\, \lambda - \iota_A\, \xi)
\ee
for vector fields $A,B$ and two-forms $\lambda,\xi$ on $M$.\footnote{This is called a Vinogradov algebroid in~\cite{Bonelli2005a,Gruetzmann2014,Ritter2015}.} The
standard 2-Courant bracket \eqref{eq:StdHigherCourantBr} appears
in exceptional generalized geometry as the natural bracket which is
compatible with the commutator algebra of generalized Lie derivatives~\cite{Berman2010,Berman2011}.

One can also introduce a flux deformation by an additional term $\gamma_T$ in the Hamiltonian function, which twists the standard 2-Courant bracket by a four-form which is necessarily closed by the classical master equation. Given an M-theory three-form $C$-field on $M$, with four-form flux $G=\dd C$, canonical transformation of the Hamiltonian function $\gamma_0$ by the degree~3 function $C=\frac1{3!}\, C_{ijk}(X)\, \psi^i\,\psi^j\,\psi^k$ on $\cM$ yields the twisted Hamiltonian function
\be
\gamma_{G} \, := \, \e^{\delta_C}\gamma_0 \, = \, F_i\, \psi^i \, + \, \frac{1}{4!}\, G_{ijkl} \, \psi^i\, \psi^j\, \psi^k\, \psi^l \ ,
\ee
and it gives the twisted standard 2-Courant bracket as
\be \label{eq:StdHigherCourantBrTwisted}
[A + \lambda, B + \xi]_{{\rm {2C}},G} \, = \, [A,B] \, + \, \cL_A \xi \, - \, \cL_B \lambda \, + \, \frac 12 \, \dd (\iota_B\, \lambda - \iota_A\, \xi) \, + \, \iota_A \iota_B G \ .
\ee

One can now use the AKSZ construction to build BV quantized sigma-models in four dimensions based on degree 3 QP-manifolds, which we shall call 2-Courant sigma-models. For the standard Lie algebroid up to homotopy on $E_0=TM$ twisted by a closed four-form flux $G$, the BV bracket is
\be 
(\,\cdot\,,\,\cdot\,)_{\mathrm{BV}} = \int_{T[1]\Sigma_4}\,\dd^4\hat z \ \bigg(\,\frac\delta{\delta\mbf X^i}\,\wedge\,\frac\delta{\delta\mbf F_i}+ \frac\delta{\delta\mbf\chi_i}\,\w\,\frac\delta{\delta\mbf\psi^i}\,\bigg) \ ,
\ee
and the classical master equation $(\mbf S,\mbf S)_{\rm{BV}}=0$ is solved by the topological threebrane action
\be \label{eq:AKSZStdHigherCourantTwisted}
\bs{S}  \, = \, \int_{T[1]\Sigma_4}\,\dd^4\hat z \ \Big(\bs{F}_i \,\bs{D} \bs{X}^i \, + \, \bs{\psi}^i \,\bs{D} \bs{\chi}_i \, + \, \bs{F}_i \, \bs{\psi}^i \, + \, \frac{1}{4!}\, \mbf G_{ijkl} \, \bs{\psi}^i \,\bs{\psi}^j \,\bs{\psi}^k \,\bs{\psi}^l \Big) \ .
\ee
Integrating over $\theta^\mu$ and restricting to degree~0 fields in \eqref{eq:AKSZStdHigherCourantTwisted} recovers the classical action \eqref{eq:Phys4Brane}.

\subsection{Dimensional reduction and effective actions}
\label{sec:DimRedMeth}

In this paper we shall also derive some novel relations amongst AKSZ sigma-models in the various dimensions $d$ through a procedure of dimensional reduction. For this, we follow~\cite{Cattaneo2009} where a practical dimensional reduction method, called `Losev's trick', is employed. Let us briefly recall the main ingredients, which are rooted in the construction of effective actions in the BV formalism.

The symplectic structure $\omega$ on the target supermanifold $\cM$
induces a natural second order differential operator $\Delta$, which
in local coordinates can be given as
\be 
\Delta=\frac12\,\omega^{\hat\imath\hat\jmath}\,
\frac{\rd{}^2}{\partial\hat X^{\hat\imath}\, \partial\hat X^{\hat\jmath}} \ ,
\ee
where $\omega^{\hat\imath\hat\jmath}$ is the inverse of
$\omega_{\hat\imath\hat\jmath}$. This pulls back to give the BV
operator $\mbf\Delta$ for the BV bracket $(\,\cdot\,,\,\cdot\,)_{\mathrm{BV}}$ on the space of AKSZ fields $\mbf\cM$; it has ghost degree~1 and satisfies $\mbf\Delta^2=0$. Since $\mbf\Delta\mbf S=0$, the AKSZ action $\mbf S$ satisfies the BV quantum master equation $\mbf\Delta\e^{-\bs{S}/\hbar}=0$ on $\mbf\cM$, which is equivalent to
\be 
\frac12\, (\mbf S,\mbf S)_{\mathrm{BV}} = \hbar \, \mbf\Delta\mbf S \ ,
\ee
and follows from nilpotency of the quantum version of the
cohomological vector field $\mbf Q-\hbar\, \mbf\Delta$. This ensures
independence of the BRST-invariant quantum field theory on the choice of
gauge fixing, provided we define the path integral
by additionally equiping $\mbf\cM$ with
a measure $\mbf\mu$ which is compatible with $\mbf\omega$~\cite{Batalin1981}.

Borrowing standard terminology from renormalization of quantum field
theory, let us now assume that the space of AKSZ fields can be
decomposed into a direct product 
\be \label{eq:MUVIR}
\mbf\cM=\mbf\cM_{\mathrm{UV}}\times\mbf\cM_{\mathrm{IR}}
\ee
of ultraviolet (UV) and infrared (IR) degrees of freedom, with a compatible decomposition of the canonical symplectic form $\mbf\omega=\mbf\omega_{\mathrm{UV}} + \mbf\omega_{\mathrm{IR}}$, where $\mbf\omega_{\mathrm{UV}}$ is a BV symplectic structure on $\mbf\cM_{\mathrm{UV}}$ and $\mbf\omega_{\mathrm{IR}}$ is a BV symplectic structure on $\mbf\cM_{\mathrm{IR}}$. Then the BV  Laplacian also decomposes as $\mbf\Delta=\mbf\Delta_{\mathrm{UV}} + \mbf\Delta_{\mathrm{IR}}$. One now `integrates out' the ultraviolet degrees of freedom to get an effective action. The integration requires a gauge fixing on the ultraviolet sector $\mbf\cM_{\mathrm{UV}}$ of the space of superfields, which means a choice of a Lagrangian submanifold $\mbf\cL \hookrightarrow \mbf\cM_{\mathrm{UV}}$. Then the effective BV action $\bs{S}_{\mathrm{eff}}$ in the infrared sector is defined as
\be
\e^{-\bs{S}_{\mathrm{eff}}/\hbar} \, := \, \int_{\mbf\cL}\, \sqrt{\mbf\mu}_{\mbf\cL} \ \e^{-\bs{S}/\hbar} \  ,
\ee
where $\sqrt{\mbf\mu}_{\mbf\cL}$ is the measure on $\mbf\cL$ induced by $\mbf\mu$. Then the effective action satisfies the quantum master equation $\mbf\Delta_{\mathrm{IR}}\e^{-\bs{S}_{\mathrm{eff}}/\hbar}=0$. A change of gauge fixing in the ultraviolet sector, corresponding to a deformation of the Lagrangian submanifold $\mbf\cL$, only changes $\e^{-\bs{S}_{\mathrm{eff}}/\hbar}$ by a $\mbf\Delta_{\rm{IR}}$-exact term. Similarly, the value of the partition function is independent of the particular choice of splitting \eqref{eq:MUVIR} by the Batalin-Vilkovisky theorem~\cite{Batalin1981}.

Formally, this technique defines a pushforward by the projection map
$\mbf\cM\to\mbf\cM_{\mathrm{IR}}$ onto the infrared sector of the
space of superfields. In the following we use this method to reduce
our AKSZ sigma-model actions to AKSZ theories in lower dimensions.

\subsection{Double dimensional reduction of twisted sigma-models}
\label{sec:DimRedflux}

As a simple example of the dimensional reduction technique described
in \S\ref{sec:DimRedMeth} above, we describe the reduction of the AKSZ threebrane sigma-model to the AKSZ membrane sigma-model with flux deformations. To motivate the reduction of threebrane flux to membrane flux, consider the simple topological threebrane action \eqref{eq:TopHFluxTerm} 
given by the pullback of a closed four-form flux $G$ on a $d$-dimensional manifold $M$ by the worldvolume map $X:\Sigma_4 \rightarrow M$. We perform a double dimensional reduction on a circle taking both the worldvolume and the target to be product manifolds $\Sigma_4=\Sigma_3\times {S}^1$ and $M=\widetilde{M}\times {S}^1$, with $\widetilde{M}$ a manifold of dimension $d-1$, and wrap the ${S}^1$ of the worldvolume around the ${S}^1$ of the target space; in other words, we regard the membranes as modes of threebranes wrapping $S^1$. We write the local coordinates on the worldvolume $\Sigma_4$ as $\sigma=(\wt{\sigma}, t)$, where $\wt{\sigma}\in\Sigma_3$ and $t$ is the coordinate on ${S}^1$. The target space coordinate indices are $I=(i,d)$, where $i=1,\ldots,d-1$ label directions along~$\widetilde{M}$. 

Wrapping the target circle means that the map $X$ has the local expression 
\be  \label{eq:DoubleDimRedXwrapping}
X \, = \, \big(X^I(\sigma)\big)\, = \, \big(\wt{X}^{i}(\wt{\sigma}), w \, t \big)
\ee
with the reduced map $\wt{X}:\Sigma_3 \rightarrow \widetilde{M}$ and $X^d=w \, t$, where $w$ is a winding number. The dimensional reduction of the action $I_{\Sigma_4,G}$ from \eqref{eq:TopHFluxTerm} is then given by $I_{\Sigma_3,H}$ from \eqref{eq:3H}, 
where the closed three-form flux $H$ on $\widetilde{M}$ is given by
\be \label{eq:DimRedHFlux}
H_{ijk} (\wt{X}) \, = \, w\, \int_{{S}^1}\, \dd t \  G_{ijkd}(\wt{X}, t) \ .
\ee 
Hence the threebrane flux $G$ reduces to a membrane flux $H$ under
double dimensional reduction on a circle. We shall now show that this
reduction also works at the level of the full AKSZ sigma-models.

We start with the $G$-twisted standard 2-Courant sigma-model
given by \eqref{eq:AKSZStdHigherCourantTwisted}, and use the
dimensional reduction method of \S\ref{sec:DimRedMeth}. We write the
expansion of an arbitrary superfield $\bs{\phi}\in\mbf\cM$ with
respect to the coordinate direction $t$ as
\be \label{eq:DimRedNotation}
\bs{\phi} \, = \, \wt{\bs{\phi}} \, + \, \bs{\phi}_t \, \theta^t \  ,
\ee
where neither $\wt{\bs{\phi}}$ nor $\bs{\phi}_t$ contain the odd coordinate $\theta^t$. If $\bs{\phi}$ has ghost number $n$, then $\wt{\bs{\phi}}$ has ghost number $n$ and $\bs{\phi}_t$ has ghost number $n-1$. We choose the infrared fields to be $(\bs{F}_t)_i$, $\wt{\bs{X}}{}^i$, $\wt{\bs{\psi}}{}^i$ and $(\bs{\chi}_{t})_i$. On the ultraviolet fields we fix the gauge by choosing the Lagrangian submanifold $\mbf\cL$ defined by
\be
\bs{X}^I_t \, = \, 0 \ , \qquad \wt{\bs{X}}{}^d \, = \, - \, w \, t \  , \qquad \bs{\psi}^i_t \, = \, 0 \ , \qquad \bs{\psi}^{d}_t \, = \, w \qquad \mbox{and} \qquad  \wt{\bs{\psi}}{}^{d} \, = \, 0 \  .
\ee
The equations of motion for $\wt{\bs{F}}_i$ and $\wt{\bs{\chi}}_i$ give $\partial_t \wt{\bs{X}}{}^i=0$ and $\partial_t \wt{\bs{\psi}}{}^i=0$, and in this way we get the AKSZ action of the $H$-twisted standard Courant sigma-model \eqref{eq:AKSZactionStdCourant} and its BV symplectic form \eqref{eq:StdCourantBVomega}
with the definitions of the fields
\be
\bs{X}^i \, = \,  \wt{\bs{X}}{}^i \  , \qquad \bs{F}_i \, = \, \int_{{S}^1}\, \dd t \  (\bs{F}_{t})_i \  , \qquad  \bs{\psi}^i \, = \, \wt{\bs{\psi}}{}^i  \qquad \mbox{and} \qquad \bs{\chi}_i \, = \, \int_{{S}^1}\, \dd t \  (\bs{\chi}_{t})_i \  ,
\ee
and $H$-flux as in \eqref{eq:DimRedHFlux}. We refer to this type of gauge fixing as a double dimensional reduction on a circle. 

It is worth stressing that the kinetic terms are necessary in this
construction because without the term $\wt{\bs{F}}_d\, \partial_t
\wt{\bs{X}}{}^d$, the term $\wt{\bs{F}}_d\, \bs{\psi}_t^d$ gives $w\,
\wt{\bs{F}}_d$, which would yield $w=0$ and vanishing $H$-flux
on-shell. An interesting feature here is that the term
coming from the Liouville potential $\psi^i\, \dd \chi_i$ of the
threebrane has been reversed via the reduction to the Liouville
potential $-\chi_i\, \dd \psi^i$ of the membrane. 
Note also that this dimensional reduction can be done at the purely bosonic level without the ghost fields: Starting from \eqref{eq:Phys4Brane}, we use the expression \eqref{eq:DoubleDimRedXwrapping} for the wrapping of $X$, and then the equations of motion for the three-form field $F_I$ and reduced two-form fields gives the bosonic part of the standard Courant sigma-model with $H$-flux in \eqref{eq:Membr1}.

By a direct computation in local coordinates, it is further possible to show that the standard 2-Courant bracket
\eqref{eq:StdHigherCourantBr} on $M=\widetilde{M}\times S^1$ suitably
reduces to the standard Courant bracket \eqref{eq:CourantBr} on
$\widetilde{M}$. The dimensional reduction of the 2-Courant
sigma-model to the Courant sigma-model is analogous to the reduction
discussed by~\cite{Berman2011} in the context of $SL(5)$ exceptional
field theory, wherein the $SL(5)$ generalized Courant bracket reduces to the $O(3,3)$ generalized Courant bracket
(C-bracket) of double field theory.

\section{AKSZ theories of topological membranes on $G_2$-manifolds}
\label{sec:AKSZmembrG2}

In this section we start to focus our attention on topological
membrane models in M-theory which have reductions to the A-model. We
study two topological membrane theories on $G_2$-manifolds. We begin
by reviewing the topological membrane model of~\cite{MQAni2005} which
is based on the Mathai-Quillen formalism,\footnote{For further details about the Mathai-Quillen formalism in general, see e.g.~\cite{MQWu1995,MQBlau1995}.} as well as the BRST model
of~\cite{Bonelli2005b}. We supplement the Mathai-Quillen construction
with an auxiliary field, analogously to the construction of~\cite{Bonelli2005b}, and we give AKSZ formulations which reproduce both membrane models after gauge fixing. 

\subsection{Topological membrane theories}
\label{sec:MQmembrAuxF}

{\underline{\sl Mathai-Quillen membrane sigma-model.} \ } Let us begin
by reviewing the topological membrane theory of~\cite{MQAni2005},
which we call the Mathai-Quillen membrane sigma-model. Let $M_7$ be an oriented
seven-dimensional Riemannian manifold with $G_2$-structure, which is
equivalent to equiping $M_7$ with a global three-form $\Phi$ that is
closed, $\dd \Phi=0$, and coclosed, $\dd \ast\Phi=0$, where $\ast$ is
the Hodge duality operator with respect to the metric $g$ of
$M_7$. Given an embedding map $X: \Sigma_3 \rightarrow M_7$, let us
introduce a local section of the cotangent bundle $T^* M_7$ by
\be
\Xi_I \, = \, \frac 1{3!}\, (*\Phi)_{IJKL}\, \partial_\mu X^J \, \partial_\nu X^K \, \partial_\rho X^L \, \epsilon^{\mu\nu\rho} \  ,
\ee
where Greek indices label local coordinates $\sigma^\mu$ on the
worldvolume $\Sigma_3$, with
$\partial_\mu:=\frac\partial{\partial\sigma^\mu}$, and capital Latin
indices label coordinates $X^I$ on $M_7$, with
$\partial_I:=\frac\partial{\partial X^I}$. The symbol
$\epsilon^{\mu\nu\rho}$ is the Levi-Civita tensor density on
$\Sigma_3$. If $\Xi_I$ vanishes, then $X(\Sigma_3)\subset M_7$ is called an associative three-cycle. 

We further introduce a ghost field $\psi^I$ on
$\Sigma_3$ with ghost number~1 and an antighost field $\chi^I$ on
$\Sigma_3$ with ghost number~$-1$. Then the action of the Mathai-Quillen membrane sigma-model is
\be \label{eq:IG2}
I_{\rm{MQ}} \, = \, \int_{\Sigma_3}\,\dd^3 \sigma \ \Big(\, \frac 12 \,g^{IJ}\, \Xi_I \, \Xi_J \, + \, \ii \chi^I \, \big(\delta \Xi_I - {\Gamma^K}_{IJ}\, \psi^J\, \Xi_K\big) \, - \, \frac 14 \, R_{IJKL}\, \psi^I \, \psi^J \, \chi^K \,\chi^L \Big) \ , 
\ee
where
\be
\delta \Xi_I - {\Gamma^K}_{IJ}\, \psi^J\, \Xi_K \, = \, \frac 12\, (*\Phi)_{IJKL}\, \nabla_\mu \psi^J \, \partial_\nu X^K \, \partial_\rho X^L\, \epsilon^{\mu\nu\rho} \  ,
\ee
with $\nabla_\mu \psi^I= \partial_\mu \psi^I + {\Gamma^I}_{JK}\,\psi^J \, \partial_\mu X^K$ given by the Levi-Civita connection of the metric $g$ pulled back to $\Sigma_3$ by $X$, and $R^I{}_{JKL}$ are the components of the Riemann curvature tensor of $g$.\footnote{Capital Latin indices are raised and lowered with the metric $g$.}
The action \eqref{eq:IG2} is invariant under the BRST transformations
\be
\delta X^I = \psi^I \ , \qquad  \delta \psi^I = 0 \qquad \mbox{and} \qquad \delta \chi^I = \ii g^{IJ}\, \Xi_J - {\Gamma^I}_{JK}\, \psi^J \, \chi^K  \ ,
\ee
which is nilpotent only on-shell, and it is BRST-exact up to the equations of motion:
\be
I_{\rm{MQ}} \, = \, \delta \Psi'_{\rm{MQ}} \qquad \mbox{with} \quad \Psi'_{\rm{MQ}} \, = \, - \frac{\ii}{2} \, \int_{\Sigma_3}\, \dd^3 \sigma\ \chi^I\, \Xi_I \  .
\ee
The fixed point locus of the BRST charge is the space of associative three-cycles~$X:\Sigma_3\to M_7$, which are membrane instantons.

Let us now linearize the BRST transformations by supplementing the Mathai-Quillen membrane sigma-model with an auxiliary field. We define an auxiliary field $b^I$ with the new BRST transformations
\be \label{eq:BRSTMQ}
\delta X^I = \psi^I \ , \qquad \delta \psi^I = 0 \  , \qquad \delta \chi^I = b^I  \qquad \mbox{and} \qquad \delta b^I = 0 \ ,
\ee
which is now nilpotent off-shell, and the membrane action is BRST-exact with the gauge fixing fermion
\be \label{eq:PsiMQ}
{\Psi}_{\rm{MQ}} \, = \, - \int_{\Sigma_3}\,\dd^3\sigma \ \chi^I\, \Big( \ii \Xi_I \, + \, \frac 12\, \Gamma_{IJK}\, \chi^J\, \psi^K \, - \, \frac 12\, g_{IJ}\, b^J \Big) \ .
\ee
Then the membrane action ${S}_{\rm{MQ}} = \delta {\Psi}_{\rm{MQ}}$ is given by
\be \label{eq:SMQ}
{S}_{ \rm{MQ}} \, = \, \int_{\Sigma_3}\,\dd^3 \sigma\ \Big( - \ii b^I\, \Xi_I + \chi^I\, \big( \ii \delta \Xi_I + \Gamma_{JIK}\,b^J\, \psi^K \big) + \frac 12\, \partial_L \Gamma_{IJK}\, \chi^I\, \chi^J\, \psi^K\, \psi^L + \frac 12\, g_{IJ}\, b^I\, b^J  \Big) \ .
\ee
The equation of motion for $b^I$ gives 
\be
b^I \, = \, \ii g^{IJ}\, \Xi_J \, - \, {\Gamma^I}_{JK}\, \chi^J\, \psi^K \ .  
\ee
Using this expression one can show that the membrane action \eqref{eq:SMQ} reduces to the Mathai-Quillen membrane action \eqref{eq:IG2}.

\medskip

{\underline{\sl BTZ membrane sigma-model.} \ } 
In \cite{Bonelli2005b} a different topological membrane action on $G_2$-manifolds is given, which is based on BRST quantization of the topological action $I_{\Sigma_3,\Phi} = \int_{\Sigma_3}\, X^*(\Phi)$; we call it the Bonelli-Tanzini-Zabzine (BTZ for short) membrane sigma-model.
With the same fields and notation as above, the action is
\be \label{eq:BTZaction}
S_{\rm{BTZ}} \, = \, - \, I_{\Sigma_3,\Phi} \, + \, \delta \Psi_{\rm{BTZ}} \  ,
\ee
with the gauge fixing fermion
\be \label{eq:PsiBTZ}
\Psi_{\rm{BTZ}} \, = \, \int_{\Sigma_3}\, \dd^3 \sigma \  \chi^I\, \Big(g_{IJ}\, \dot{X}^J \, + \, \Phi_{IJK}\, \partial_1 X^J\, \partial_2 X^K \, + \, \frac 12\, \Gamma_{IJK}\,\chi^J\, \psi^K \, - \, \frac 12\, g_{IJ}\, b^J \Big) \  ,
\ee
where the worldvolume indices run through $\mu=0,1,2$ and the dot denotes the action of the derivative $\partial_0$. 
 The BRST transformations are the same as those of the Mathai-Quillen
 membrane model in \eqref{eq:BRSTMQ}, thus they have identical BV
 formulations. Since $\dd \Phi =0$, the topological flux term
 $I_{\Sigma_3,\Phi}$ in the AKSZ framework arises from a canonical
 transformation as in \S\ref{sec:CourantAlgAKSZ}, and
 consequently it has no effect in the BV algebra on the mapping space
 $\mbf
\cM$. Hence in the following we will only study the BRST-exact term in \eqref{eq:BTZaction}.

\subsection{BV formulation and AKSZ constructions}
\label{sec:AKSZmembr1}

Both topological membrane sigma-models are described by a gauge fixing fermion $\Psi\big[X^I,\psi^I, \chi^I, b^I\big]$. The only non-zero BRST transformations are $\delta X^I=\psi^I$ and $\delta \chi^I = b^I$, so 
\be
\delta\Psi \, = \, \int_{\Sigma_3}\, \dd^3 \sigma \  \Big( \psi^I\, \frac{\delta \Psi}{\delta X^I} \, + \, b^I\, \frac{\rdelta \Psi}{\delta \chi^I}\Big) \  .
\ee
With the definition of the antifields\footnote{As in \S\ref{sec:gaugefixing} we denote the antifield of a field $\phi$ by $\phi^+$.}
\be
X^+_I \, = \, \frac{\delta \Psi}{\delta X^I} \qquad \mbox{and} \qquad \chi^+_I \, = \, \frac{\rdelta \Psi}{\delta \chi^I} \ , 
\ee
we can rewrite the BRST-exact part of the membrane actions as
\be \label{eq:SG2}
\delta\Psi\, = \, \int_{\Sigma_3}\, \dd^3 \sigma \ \big( \psi^I\, X^+_I \, + \, b^I\, \chi^+_I \big) \ .
\ee
Thus the BRST-exact membrane actions in \eqref{eq:SMQ} and
\eqref{eq:BTZaction} differ only in the choice of gauge fixing,
i.e. in the choice of Lagrangian submanifold $\mbf\cL\subset\mbf\cM$. In the following we propose two different AKSZ constructions for these topological membrane theories.

\medskip

{\underline{\sl AKSZ construction I.} \ }  
Our first AKSZ construction contains a rather large number of fields, but very few of them are explicitly used in the gauge fixed action. 
The source dg-manifold is $ \cW=T [1]\Sigma_3$ as usual, and the target symplectic dg-manifold is $\cM={ T }^*[2] T [-1] T [1]M_7$. The base coordinates in $ T [-1] T [1]M_7$ are $(X^I,\xi^I,B^I,\eta^I)$ with degree $(0,1,0,-1)$, where $X^I$ are associated to the coordinates of $M_7$. The graded fiber coordinates are $(F_I,\zeta_I, N_I, G_I)$ with degree $(2,1,2,3)$, and the canonical symplectic structure of degree~2 on $\cM$ is
\be
\omega_{3,\rm{I}} \, = \, \dd F_I \w \dd X^I  \, + \, \dd \zeta_I \w \dd \xi^I  \, + \, \dd {N}_I \w \dd B^I \, + \, \dd G_I \w \dd \eta^I \ .
\ee
In the following we expand a general AKSZ superfield
$\mbf\phi\in\mbf\cM$ as in \eqref{eq:superfieldexp} for $d=3$.
Our membrane BRST fields $X^I,\psi^I,\chi^I,b^I$ do not have form components, so we choose them as the zeroth or third components of a superfield. Our choice in this first construction is as the zeroth component for both membrane models, and their antifields are assigned to the third components. Explicitly this means we take
\be\begin{aligned}
X^{(0)\, I} & =X^I & \quad &\mbox{and}& \quad F^{(3)}_I & =X^{+}_I  \ , \\[4pt]
\xi^{(0)\, I} & =   \, \psi^I  & \quad &\mbox{and}& \quad \zeta^{(3)}_I & =   \, \psi^{+}_I \ , \\[4pt] 
\eta^{(0)\, I} & = \chi^I   & \quad &\mbox{and}& \quad G^{(3)}_I & =\chi^{+}_I \ , \\[4pt] 
B^{(0)\, I} & =   \, b^I  & \quad &\mbox{and}& \quad {N}^{(3)}_I & =   \, b^+_I \  . \end{aligned} \ee

The AKSZ action is constructed without kinetic terms and with a
degree~3 Hamiltonian function $\gamma$ such that the corresponding BV
bracket with the associated cohomological vector field $\mbf Q$ on
$\mbf\cM$ generates the BRST transformations \eqref{eq:BRSTMQ}. Thus we take
\be \label{eq:AKSZmembr1}
\bs{S}_{G_2, \rm{I}} \, = \, \int_{ T [1]\Sigma_3}\,\dd^3\hat z \ \big(  \bs{\xi}^I\, \bs{F}_I \, +  \, \bs{B}^I\, \bs{G}_I \big) \  ,
\ee
which has eight components after the expanding the superfields. We use a gauge fixing fermion to set the antifields $X^{+}_I$, $\psi^{+}_I$, $\chi^{+}_I$,  $b^+_I$, and we choose the gauge fixing of \eqref{eq:AKSZmembr1} on the other fields to give the gauge fixed action \eqref{eq:SG2}.
For example, we may choose the Lagrangian submanifold $\mbf\cL$ determined by
the equations
\be\begin{aligned} 
X^{(1)\, I}=X^{(3)\, I} &=0 & \quad &\mbox{and}& \qquad F^{(1)}_I&=0 \ , \\[4pt] 
\xi^{(1)\, I}=\xi^{(3)\, I}&=0 & \qquad &\mbox{and}& \qquad \zeta^{(1)}_I&=0 \ , \\[4pt]
\eta^{(1)\, I}=\eta^{(3)\, I}&=0 & \qquad &\mbox{and}& \qquad G^{(1)}_I&=0 \ , \\[4pt]
B^{(1)\, I}=B^{(3)\, I}&=0 & \quad &\mbox{and}& \qquad {N}^{(1)}_I&=0 \ ,
\end{aligned} \ee
for the antifields.
The other antifields given by the gauge fixing fermion are $X^+_I$, $\chi^+_I$, $\psi^+_I$ and $b^+_I$. If we choose \eqref{eq:PsiMQ} we get the Mathai-Quillen membrane action \eqref{eq:SMQ}, while if we choose \eqref{eq:PsiBTZ} we get the BRST-exact part of the BTZ topological membrane action~\eqref{eq:BTZaction}.

For example, in the Mathai-Quillen membrane sigma-model the pertinent antifields are given by
\be \begin{aligned}
  X^+_I  &= \,  \frac{\delta{\Psi}_{\rm{MQ}}}{\delta X^I}  = \, -\, \ii\frac{\delta}{\delta X^I}\int_{\Sigma_3}\,\dd^3 \sigma \ \chi^I\, \Xi_I \, - \frac 12\, \partial_I \Gamma_{JKL}\,\chi^J\, \chi^K\, \chi^L \, + \, \frac 12\, \partial_I g_{JK}\, \chi^J\, b^K \ , \\[4pt]
\chi^+_I  &= \,  \frac{\rdelta{\Psi}_{\rm{MQ}}}{\delta \chi^I}  = \,-\, \ii \Xi_I \, - \, \Gamma_{[IJ]K}\, \chi^J\, \psi^K \, + \, \frac 12\, g_{IJ}\, b^J \, ,
\end{aligned}\ee
and it is easy to see
\be
\psi^I\, \frac{\delta}{\delta X^I} \int_{\Sigma_3}\,\dd^3 \sigma \ \chi^I\, \Xi_I \, = \, - \chi^I\, \delta\Xi_I \ ,
\ee
so that gauge fixing the antifields in this way restricts the AKSZ
action functional \eqref{eq:AKSZmembr1} on $\mbf\cL$ to the action \eqref{eq:SMQ}. The gauge fixing with $\Psi_{\rm{BTZ}}$ is very similar, and it gives the BTZ membrane action \eqref{eq:BTZaction}.
Note that it is possible to add kinetic terms to the AKSZ action, and then set them to zero with a more specific gauge fixing choice, but evidently the model \eqref{eq:AKSZmembr1} is simpler to work with.

\medskip

{\underline{\sl AKSZ construction II.} \ }  
We introduce another AKSZ construction for both topological membrane
theories, which is based on the standard Courant sigma-model from
\S\ref{sec:CourantAlgAKSZ}. The BV action that we want to reproduce in
the AKSZ theory is again \eqref{eq:SG2}, but now we define the
fermionic fields $\psi^I$ and $\chi^I$ as one-forms in the superfield
formalism. The target in this case is taken to be the QP-manifold
$\cM= T ^*[2] T [1]M_7$ of degree~2 corresponding to the standard
Courant algebroid on $TM_7\oplus T^*M_7$, which contains half as many
coordinates compared to the previous construction. The notation for
the coordinates are the same as before, so that $(X^I, F_I, \xi^I, \zeta_I)$ have degrees $(0,2,1,1)$. The symplectic form is
\be
\omega_{3,\rm{II}} \, = \, \dd F_I \w \dd X^I  \, + \, \dd \zeta_I \w \dd \xi^I \ .
\ee
The relevant fields in the superfield formalism are
\be\begin{aligned}
X^{(0)\, I} & = X^I & \quad &\mbox{and}& \quad X^{(1)\, I}_{0} & = \chi^I \ , \\[4pt]
\big(F^{(2)}_{I}\big)_{12}  & = \chi^+_I  & \quad &\mbox{and}& \quad \big(F^{(3)}_{I}\big)_{012} & = X^+_I \ , \\[4pt] 
\xi^{(0)\, I} & = \psi^I   & \quad &\mbox{and}& \quad \xi^{(1)\, I}_0 & = b^I \ , \\[4pt] 
\big(\zeta^{(2)}_{I}\big)_{12} & = -\, b^+_I & \quad &\mbox{and}& \quad \big(\zeta^{(3)}_{I}\big)_{012} & =  \psi^+_I \  ,
\end{aligned} \ee
where we used an explicit worldvolume index convention to define the membrane fields $\chi^I$, $b^I$ and their antifields.
The BV action then simply corresponds to the untwisted Hamiltonian function $\gamma_0$ from \S\ref{sec:CourantAlgAKSZ} and reads
\be \label{eq:AKSZmembr2}
\bs{S}_{G_2,\rm{II}} \, = \, \int_{ T [1]\Sigma_3}\,\dd^3\hat z \ \bs{\xi}^I\, \bs{F}_I \  .
\ee
There are many possible gauge fixings which recover the action
\eqref{eq:SG2}. One choice is to take the Lagrangian submanifold
defined by
\be\begin{aligned}
F^{(0)}_I=F^{(1)}_I & =  0 & \qquad &\mbox{and}& \qquad \zeta^{(0)}_I =\zeta^{(1)}_I& =  0  \ , \\[4pt]
\big(F^{(2)}_{I}\big)_{01 } & = 0 & \qquad &\mbox{and}& \qquad \big(\zeta^{(2)}_{I}\big)_{01 }&  =  0 \ , \\[4pt]
\big(F^{(2)}_{I}\big)_{13 } & = 0 & \qquad &\mbox{and}& \qquad \big(\zeta^{(2)}_{I}\big)_{3 }& =  0  \ .
\end{aligned}\ee
The residual antifields are again set by the gauge fixing fermion $\Psi\big[X^I,\psi^I,\chi^I,b^I\big]$, given in \eqref{eq:PsiMQ} for the Mathai-Quillen membrane sigma-model and in \eqref{eq:PsiBTZ} for the BTZ membrane sigma-model.

It is an interesting feature of our first AKSZ construction that the two terms in \eqref{eq:AKSZmembr1} are decoupled from each other, in the sense that they can be gauge fixed separately and decoupled in the AKSZ action as well. This means that one can remove the second term with a gauge fixing to get our second AKSZ constructions, but they differ from those proposed for the topological membranes, because the antifields are assigned differently.

\subsection{Derived brackets}
\label{sec:IndBracketAKSZ1}

The main geometric distinction between the two AKSZ membrane theories
we have constructed above is that the second construction is based on
a target which is a QP-manifold of degree~2, corresponding to the
standard Courant algebroid, whereas the first construction is based on
a target which is not an N-manifold, as it involves local affine
coordinates of degree $-1$, and consequently does not correspond to a
symplectic Lie 2-algebroid. Passing to dg-manifolds which are equiped
with negative gradings is of course natural and standard in the
BV--BRST formalism, wherein ghost fields and antifields typically
come with negative gradings,
but it takes us out of the realm of graded geometry into derived geometry~\cite{Toen2014}:
Whereas non-negatively graded symplectic dg-manifolds generally
correspond to symplectic $L_\infty$-algebroids, those which are
arbitrarily graded correspond to \emph{derived} symplectic
$L_\infty$-algebroids. The relevance of $L_\infty$-algebroids in BV
quantization was already emphasised by~\cite{Zwiebach1992,AKSZ1997},
but entering into further discussion of these geometric facts would take us
far away from the scope of the present paper, so we content ourselves
in pointing out a few interesting geometric consequences of the corresponding
derived bracket construction. 

The degree~3 Hamiltonian function on $\cM={ T }^*[2] T [-1] T [1]M_7$ corresponding to the first AKSZ action \eqref{eq:AKSZmembr1} is given by
\be \label{eq:HamAKSZmembr1}
\gamma_{G_2,\rm{I}}=F_I\, \xi^I+ G_I\, B^I\ .
\ee
Its first term is the same as the Hamiltonian function
\eqref{eq:HamStdCourant} for the standard Courant algebroid, so its
derived brackets gives the standard Courant bracket
\eqref{eq:CourantBr} on degree 1 functions of
$(X,\xi,\zeta)$. Moreover, this is also the derived bracket of the
Hamiltonian function corresponding to the second AKSZ action
\eqref{eq:AKSZmembr2}, which contains solely the first term of \eqref{eq:HamAKSZmembr1}.

The interesting feature here is the consequence of the second term in \eqref{eq:HamAKSZmembr1} and the negative degree coordinates $\eta^I$. The derived bracket of a symplectic dg-manifold with symplectic structure of degree~2 is defined on degree 1 functions. Such a function $f$ can be expanded in the form
\be
f \, = \, f^{(0)}(X,B,\xi,\zeta) \, + \, f_I^{(1)}(X,B,\xi,\zeta,F,N) \, \eta^I \, + \, \sum^7_{l=2}\, f_{I_1\cdots I_l}^{(l)}(X,B,\xi,\zeta,F,N,G) \, \eta^{I_1}\cdots \eta^{I_l} \ ,
\ee
where $f^{(l)}$ is an $l$-form in the non-negatively graded coordinates on $\cM$ of degree $l+1$. 
The second term $G_I\, B^I$ in the Hamiltonian function decouples on
the zeroth order functions $f^{(0)}(X,B,\xi,\phi)$, since it does not
contain any of the canonically conjugate coordinates to $X$, $B$,
$\xi$ or $\zeta$. Hence our derived bracket is closed on the subspace
of zeroth order functions $f^{(0)}$, where it gives the standard Courant
bracket \eqref{eq:CourantBr}, with the coefficients now depending on
the two degree~0 coordinates $X$ and $B$. The degree~0 fields are doubled in this sense, but they play an asymmetric role in the underlying geometric structure.

The restriction of the derived bracket to any higher order in $\eta^I$
is no longer closed, and only closes if we consider all orders at
once. Thus our derived bracket appears as an infinite extension of the
standard Courant bracket, which contains the standard Courant bracket
as the subalgebra of functions which are independent of $\eta^I$. This
structure underlies the derived symplectic $L_\infty$-algebroid over
$M_7$ alluded to above.\footnote{See e.g.~\cite{Gruetzmann2014} for a general definition of
  $L_\infty$-algebroids.}

\subsection{Dimensional reductions from topological threebrane theories}
\label{sec:3braneAndMembrane}

In \S\ref{sec:LieAlguth} we introduced an AKSZ topological threebrane
sigma-model which has the standard 2-Courant bracket as its
derived algebraic structure on a graded target space which is a
QP-manifold of degree~3. We can shed further light on the algebroid
structure discussed in \S\ref{sec:IndBracketAKSZ1} by considering our
membrane models as arising through certain reductions of such a threebrane
theory. We first consider this sigma-model without a
four-form flux deformation and defined for the $G_2$-manifold $M=M_7$. We suppose that the threebrane worldvolume is a product manifold $\Sigma_4=\Sigma_3 \times S^1$, and that all superfields are independent of the extra coordinate $t$ of $S^1$. Using the same notation \eqref{eq:DimRedNotation} for the expansion of an arbitrary superfield, integration over the odd coordinate $\theta^t$ in the action \eqref{eq:AKSZStdHigherCourantTwisted} without the flux term leads to the AKSZ action\footnote{The kinetic part of the AKSZ action is given here by $-\bs{\vartheta}$, where $\vartheta$ is the Liouville potential on the symplectic dg-manifold $\cM$.}
\be
\bs{S}_{3,\rm{red}} \, = \, \int_{ T [1]\Sigma_3}\,\dd^3\hat z \ \big( \bs{G}_I\, \bs{B}^I \, + \, \bs{F}_I\, \bs{\xi}^I \, - \,\bs{F}_I\, \bs{D}\mbf X^I \, + \, \bs{\xi}^I\, \bs{D}  \bs{\zeta}_I \, - \, \bs{ G}_I\, \bs{D} \bs{\eta}^I\, - \, \bs{B}^I\, \bs{D} \bs{N}_I \big)
\ee 
and the BV symplectic form
\be
{\mbf\omega}_{3,\rm{red}} \, = \, \int_{ T [1]\Sigma_3}\,\dd^3\hat z \ \big(\bs{\delta} \bs{F}_I\, \bs{\delta} \bs{X}^I  \, - \, \bs{\delta} \bs{\zeta}_I\, \bs{\delta} \bs{\xi}^I \, + \, \bs{\delta} \bs{G}_I\, \bs{\delta} \bs{\eta}^I \, - \, \bs{\delta} \bs{N}_I\, \bs{\delta} \bs{B}^I \big) \  ,
\ee
where we have introduced the fields
\be \begin{aligned}
{\bs{F}}_I \, &= \, - \, (\bs{ F}_{t})_{ I} \ , \qquad & {\bs{G}}_I \, &= \,\wt{\bs{ F}}_I\ , \qquad & {\bs{\zeta}}_I \, &= \, (\bs{\chi}_{t})_{ I} & \qquad &\mbox{and} & \qquad \bs{N}_I \, &= \, \wt{\bs{\chi}}_I \ , \\[4pt]
{\bs{X}}^I \, &= \, {\wt{\bs{X}}}{}^I \ , \qquad & {\bs{\eta}}^I \, &= \, \bs{X}^I_t \  , \qquad & {\bs{B}}^I \, &= \, - \, \bs{\psi}^I_{t} & \qquad & \mbox{and} & \qquad {\bs{\xi}}^I \, &= \, - \, {\wt{\bs{\psi}}}{}^I \  ,  
\end{aligned}\ee
and rescaled them by the length of $S^1$.
Thus the reduced AKSZ action without the kinetic terms is our first AKSZ membrane action \eqref{eq:AKSZmembr1}, up to a few sign differences appearing in the symplectic forms which can be resolved with a redefinition of the original symplectic form of the membrane sigma-model that leaves its gauge fixed action invariant. On the other hand, the kinetic terms can be removed with the same gauge fixing that we used to obtain the topological membrane theories in this section. In this way, the threebrane AKSZ action without any kinetic term
\be \label{eq:AKSZ3branePure}
 \int_{ T[1] {\Sigma_4}}\,\dd^4\hat z \ \bs{ F}_I\, \bs{\psi}^I
 \ee
is a straightforward extension of our AKSZ membrane sigma-models.

This means therefore that our first AKSZ construction for topological membranes on $G_2$-manifolds is a reduced AKSZ theory of topological threebranes on the same target space. The special feature of the threebrane theory is that its derived bracket on the target QP-manifold $T^*[3]T[1]M_7$ of degree~3 gives the standard 2-Courant bracket \eqref{eq:StdHigherCourantBr} on the vector bundle $ E= T M_7\oplus \bigwedge^2\,  T ^*M_7$, which relates the geometry behind our specific AKSZ construction to the exceptional generalized geometry of M-theory.

The second AKSZ construction for topological membranes from \S\ref{sec:AKSZmembr1} can also be reformulated within a topological threebrane sigma-model, in the same way as the first construction. The only difference is that we get an additional term in the AKSZ action after the reduction, which can be set to zero with gauge fixing, because we do not need those fields to get the topological membrane theories with further gauge fixing. Hence the action \eqref{eq:AKSZ3branePure} reduces to the second AKSZ sigma-model action as well. 

In \S\ref{sec:DimRedflux} we saw that viewing membranes as wrapping modes of threebranes, by wrapping the worldvolume circle on the target circle, reduces the four-dimensional standard 2-Courant sigma-model with $G$-flux to the three-dimensional standard Courant sigma-model with $H$-flux. This means that it is possible to add $G$-flux to our topological membrane theories at the threebrane level. Although the reduction above, wherein the fields are taken to be independent of one worldvolume direction, removes the topological flux term in \eqref{eq:TopHFluxTerm}, at the level of the full AKSZ action it does not. It leaves an extra contribution
\be\label{eq:extrafluxterm}
\frac{1}{3!}\, \int_{{T}[1]\Sigma_3}\,\dd^3\hat z \ \mbf G_{IJKL}\, \bs{\xi}^I\, \bs{\xi}^J\, \bs{\xi}^K\, \bs{B}^L \ ,
\ee
which can be taken as a definition of a flux deformation for our first AKSZ membrane construction in~\S\ref{sec:AKSZmembr1}.

Alternatively, one can directly induce the topological flux deformation $I_{\Sigma_3,\Phi}$ that we neglected in the action \eqref{eq:BTZaction} by applying the double dimensional reduction technique from \S\ref{sec:DimRedflux}. For this, we first note that, generally, the AKSZ threebrane sigma-model \eqref{eq:AKSZStdHigherCourantTwisted} gives the BV action for the sigma-model of~\cite{Bonelli2005a} for topological threebranes on an eight-dimensional $Spin(7)$-manifold $M_8$, with the twist $G$ taken to be the global self-dual closed four-form corresponding to the $Spin(7)$-structure on $M_8$~\cite{Ikeda2011}, whose path integral localizes on Cayley four-cycles (threebrane instantons). We can then embed our topological brane sigma-models with target $G_2$-manifold $(M_7,\Phi)$ into this threebrane theory by taking $\Sigma_4=\Sigma_3\times S^1$ and $M_8=M_7\times S^1$ with the Cayley four-form
\be 
G = \dd X^8\w\Phi+\ast\Phi \ .
\ee
Using double dimensional reduction on a circle as in \S\ref{sec:DimRedflux} then reproduces the $H$-twisted standard Courant sigma-model \eqref{eq:AKSZactionStdCourant} with flux $H=w\,\Phi$, and consequently leads to our second AKSZ construction from~\S\ref{sec:AKSZmembr1} with topological term. On the other hand, if the original threebrane is localized on $S^1$, i.e. $X^8$ is constant, then the threebrane theory reduces on $t$-independent superfields as above to our first AKSZ construction, with extra flux term \eqref{eq:extrafluxterm} given by $G=\ast\Phi$. In this setting these threebrane worldvolume theories are regarded as providing a microscopic description of topological F-theory~\cite{Anguelova2004,Bonelli2005b}.

\section{AKSZ theories for the topological A-model}
\label{sec:AKSZtopAmod}

The A- and B-models of the topological sigma-model~\cite{Witten1988} have been extensively studied over the past three decades, particularly when they are coupled to gravity where they become the A- and B-models of topological string theory. They were also one of the first examples of the AKSZ construction from~\cite{AKSZ1997}. In particular, all known AKSZ constructions for the A-model are Poisson sigma-models~\cite{AKSZ1997,Park2000,Ikeda2012,Pestun2006,Zucchini2004,Stojevic2005}, so they all have the same target QP-manifold of degree~1, symplectic structure and Hamiltonian function as in \S\ref{sec:AKSZPoisson}. The Poisson bivector $\pi$ in these instances is given by the inverse of the K{\"a}hler form on the target Calabi-Yau manifold, and the AKSZ sigma-models all reduce to the A-model in particular gauges. 

In this section we will follow the general procedure of \S\ref{sec:DimRedMeth} to compute a dimensional reduction, at the level of the AKSZ construction, for both AKSZ topological membrane theories which we derived in \S\ref{sec:AKSZmembr1}. In each case the reduction leads to a new AKSZ formulation for the topological A-model which differs from the Poisson sigma-model.

\subsection{Dimensional reduction of AKSZ membrane sigma-models}
\label{sec:DimRedAKSZmembr}

We begin by applying a canonical transformation as described in \S\ref{sec:AKSZingeneral}. Here we will only use infinitesimal canonical transformations, which act on functions $\bs{f}$ on the phase space $\mbf\cM$ as
\be \label{eq:inftezCanTrafo}
\bs{f} \, \longmapsto \, {}^{\mbf\alpha}\bs{f} \, = \, \bs{f} \, + \, \varepsilon\,\bv{\bs{f}}{\bs{\alpha}} \  ,
\ee  
where $\varepsilon$ is an infinitesimal parameter and $\bs{\alpha}$ is a fermionic functional of the fields with ghost number~$-1$.
We perform such a canonical transformation on our two AKSZ membrane actions to induce kinetic terms, which will be used for dimensional reduction.

For the first AKSZ membrane action \eqref{eq:AKSZmembr1}, the fermionic functional we choose is
\be
\bs{\alpha} \, = \, \int_{ T [1]\Sigma_3}\,\dd^3\hat z \ \big(\bs{\zeta}_I\, \bs{D} \bs{X}^I \, + \, \bs{N}_I\, \bs{D} \bs{\eta}^I \big) \ ,
\ee
where as previously the superworldvolume differential is
$\bs{D}=\theta^\mu\, \partial_\mu$. Calculating the BV bracket $\bv{\bs{S}_{G_2,\rm{I}}}{\bs{\alpha}}$ term by term we get the BRST-equivalent action
\be\begin{aligned} \label{eq:AKSZmembr1CT}
{}^{\mbf\alpha}\bs{S}_{G_2,\rm{I}} \, &= \, \bs{S}_{G_2, \rm{I}} \, + \, \varepsilon\,\bv{\bs{ S}_{G_2,\rm{I}}}{\bs{\alpha}} \\[4pt]
 & = \, \int_{ T [1]\Sigma_3}\,\dd^3\hat z \ \Big(\bs{F}_I\,\bs{\xi}^I \, + \, \bs{B}^I\, \bs{G}_I \, + \, \varepsilon\,\big(\bs{F}_I\,\bs{D} \bs{X}^I \, + \, \bs{\xi}^I\, \bs{D} \bs{\zeta}_I \, + \,\bs{B}^I\, \bs{D}{\bs{N}}_I \, - \, \bs{G}_I\, \bs{D} \bs{\eta}^I \big)\Big) \  .
\end{aligned}\ee
Similar considerations apply to the second action
\eqref{eq:AKSZmembr2}: If we restrict the functionals and hence also
the action to half of the fields $\bs{F}$, $\bs{X}$, $\bs{\xi}$ and
$\bs{\zeta}$, we get the fermionic functional of the canonical
transformation $\mbf\alpha=\int_{ T [1]\Sigma_3}\,\dd^3\hat z \ \bs{\zeta}_I\, \bs{D} \bs{X}^I$ which gives us the BRST-equivalent action
\be
{}^{\mbf\alpha}\bs{ S}_{G_2, \rm{II}} \, = \, \int_{ T
  [1]\Sigma_3}\,\dd^3\hat z \ \Big(\bs{F}_I\,\bs{\xi}^I \, + \, \varepsilon\,\big(\bs{F}_I\,\bs{D} \bs{X}^I \, + \, \bs{\xi}^I\, \bs{D} \bs{\zeta}_I \big)\Big) \ .
\ee  

Now let us turn to the dimensional reduction of the AKSZ membrane
sigma-models. We assume that the target and worldvolume manifolds are
products $M_7=M_6\times S^1$ and $\Sigma_3=\Sigma_2 \times S^1$, where
the coordinates of the target and worldvolume circles are indexed by
$I=7$ and $\mu=t$ respectively. We use again the expansion
\eqref{eq:DimRedNotation} of an arbitrary superfield $\bs{\phi}\in \mbf\cM$.
In terms of expanded superfields, the symplectic structure is given by
\be \begin{aligned}
\bs{\omega}_{3,\rm{I}}\, = \, \int_{ T [1]\Sigma_2}\,\dd^2\hat z \ \int_{S^1}\, \dd t\
\big( & -\,\bs{\delta} \wt{\bs{F}}_I\, \bs{\delta} \bs{X}_t^I \, - \, \bs{\delta} (\bs{F}_{t})_{ I}\, \bs{\delta} \wt{\bs{X}}{}^I \, - \, \bs{\delta} \wt{\bs{\zeta}}_I\, \bs{\delta} \bs{\xi}_t^I \, + \, \bs{\delta} (\bs{\zeta}_{t})_{ I}\, \bs{\delta} \wt{\bs{\xi}}{}^I  \\
& \ - \, \bs{\delta} \wt{\bs{G}}_I\, \bs{\delta} \bs{\eta}_t^I \, + \, \bs{\delta} (\bs{G}_{t})_{I}\, \bs{\delta} \wt{\bs{\eta}}{}^I \, - \, \bs{\delta} \wt{\bs{N}}_I\, \bs{\delta} \bs{B}_t^I \, - \, \bs{\delta} ({\bs{N}}_{t})_{I}\, \bs{\delta} \wt{\bs{B}}{}^I\big) \  ,
\end{aligned}\ee
and the action \eqref{eq:AKSZmembr1CT} by
\be \begin{aligned}
{}^{\mbf\alpha}\bs{S}_{G_2, \rm{I}} \, =& \, \int_{ T
  [1]\Sigma_2}\,\dd^2\hat z \ \int_{S^1}\,\dd t\ \Big( \wt{\bs{F}}_I\, \bs{\xi}_t^I \, - \, (\bs{F}_{t})_{I}\, \wt{\bs{\xi}}{}^I \, + \, \wt{\bs{B}}{}^I\, (\bs{G}_{t})_{ I} \, - \, \bs{B}_t^I\, \wt{\bs{G}}_I \\
& + \, \varepsilon\,\big(\wt{\bs{F}}_I\, \wt{\bs{D}} \bs{X}_t^I \, + \, \wt{\bs{F}}_I\, \partial_t \wt{\bs{X}}{}^I \, - \, (\bs{F}_{t})_{ I}\, \wt{\bs{D}} \wt{\bs{X}}{}^I \, + \, \wt{\bs{\xi}}_I\, \wt{\bs{D}} \bs{\zeta}_t^I \, - \, \wt{\bs{\xi}}_I\, \partial_t \wt{\bs{\zeta}}{}^I \, + \, (\bs{\xi}_{t})_{ I}\, \wt{\bs{D}} \wt{\bs{\zeta}}{}^I \\
 & - \, \wt{\bs{G}}_I\, \wt{\bs{D}} \bs{\eta}_t^I \, + \, \wt{\bs{G}}_I\, \partial_t \wt{\bs{\eta}}^I \, - \, (\bs{G}_{t})_{ I}\, \wt{\bs{D}} \wt{\bs{\eta}}{}^I \, + \,
  \wt{\bs{B}}_I\, \wt{\bs{D}} {\bs{N}}_t^I \, + \, \wt{\bs{B}}_I\, \partial_t \wt{\bs{N}}{}^I \, - \, (\bs{B}_{t})_{ I}\, \wt{\bs{D}} \wt{\bs{N}}{}^I \big) \Big) \  .
\end{aligned}\ee

We choose $\wt{\bs{F}}$, $\bs{X}_t$, $\wt{\bs{\phi}}$, $\bs{\xi}_t$, $\wt{\bs{G}}$, $\bs{\eta}_t$, $\wt{\bs{N}}$ and $\bs{B}_t$ to be the ultraviolet fields, and the rest to be the infrared fields. We define the gauge fixing condition as the Lagrangian submanifold $\mbf\cL$ defined by $\bs{X}_t=\bs{\eta}_t=\bs{\xi}_t=\bs{B}_t=0$, and then integrate out the remaining ultraviolet fields. This leads to the conditions $\partial_t \wt{\bs{X}}=\partial_t \wt{\bs{\xi}}= \partial_t\wt{\bs{\eta}} =\partial_t \wt{\bs{B}}=0$, so these fields do not depend on $t$. We also integrate out all of the fields with $I=7$ index, and introduce new fields
\be 
{\bs{\chi}}_i \, = \, - \int_{S^1}\,\dd t \ ({\bs{F}}_{t})_{i} \  , \quad  {\bs{p}}_i \, = \, \int_{S^1}\,\dd t \  ({\bs{\zeta}}_{t})_{i}\  , \quad   {\bs{h}}_i \,  = \, \int_{S^1}\,\dd t \  ({\bs{G}}_{t})_{i}\quad \mbox{and} \quad \bs{n}_i \,  = \, - \int_{S^1}\,\dd t \  ({\bs{N}}_{t})_{i} \  , 
\ee
and
\be
{\bs{X}}^i \, = \, {\wt{\bs{X}}}{}^i \  , \qquad   {\bs{q}}^i \,  = \, {\wt{\bs{\xi}}}{}^i \ , \qquad   {\bs{\eta}}^i \, = \, {\wt{\bs{\eta}}}{}^{\,i} \qquad \mbox{and} \qquad {\bs{b}}^i \,  = \, {\wt{\bs{B}}}{}^i \  ,  
\ee
where we used the index notation $I=(i,7)$ with $i=1,\dots,6$ the
coordinate directions along $M_6$. Our effective action is then
\be \label{eq:AmodDRAction}
\bs{S}_{G_2,\rm{I}}^{\rm{eff}} \, = \, \int_{ T
  [1]\Sigma_2}\,\dd^2\hat z \ \Big({\bs{\chi}}_i\, {\bs{q}}^i \, + \, {\bs{b}}^i\, {\bs{h}}_i \, + \, \varepsilon\, \big({\bs{\chi}}_i\, \bs{D} {\bs{X}}^i \, + \, {\bs{q}}^i\, \bs{D} {\bs{p}}_i \, - \, {\bs{h}}_i\, \bs{D} {\bs{\eta}}^i \, - \, {\bs{b}}^i\, \bs{D} {\bs{n}}_i\big) \Big) \ ,
\ee
and the new symplectic form is
\be \label{eq:AKSZsymplAmodBV1}
\bs{\omega}_{2,\rm{I}} \, = \, \int_{ T [1]\Sigma_2}\,\dd^2\hat z \ \big(\bs{\delta} \bs{\chi}_i\, \bs{\delta} \bs{X}^i \, + \, \bs{\delta} \bs{p}_i\, \bs{\delta} \bs{q}^i \, + \, \bs{\delta} \bs{h}_i\, \bs{\delta} \bs{\eta}^i \, + \, \bs{\delta} {\bs{n}}_i\, \bs{\delta} \bs{b}^i \big) \ .
\ee

We now perform another infinitesimal canonical transformation with the same parameter $\varepsilon$ and the fermion
\be
\mbf\alpha' \, = \, - \int_{ T [1]\Sigma_2}\,\dd^2\hat z \ \bs{p}_i\, \bs{D} \bs{X}^i \, - \, \int_{\Sigma_2}\, \dd^2 \sigma\ \Big({n}^{(0)}_i\, \dd \eta_i^{(1)} \, - \, {n}^{(1)}_i\, \dd \eta_i^{(0)}\Big)
\ee
in order to eliminate the kinetic terms. In this way we arrive at the action
\be\label{eq:AKSZAmod1}
\bs{S}_{\rm{A}, \rm{I}} \, = \, \int_{ T [1]\Sigma_2}\,\dd^2\hat z \ \big({\bs{\chi}}_i\, {\bs{q}}^i \, + \, {\bs{b}}^i\, {\bs{h}}_i \big) \ .
\ee 
If we restrict this construction and the dimensional reduction to half
of the fields $\bs{F}$, $\bs{X}$, $\bs{\xi}$ and $\bs{\zeta}$, we
arrive at
the action for the dimensional reduction of our second AKSZ membrane
model in the form
\be\label{eq:AKSZAmod2}
\bs{S}_{\rm{A}, \rm{II}} \, = \, \int_{ T [1]\Sigma_2}\,\dd^2\hat z \ {\bs{\chi}}_i\, {\bs{q}}^i \ .
\ee

In the following we will introduce AKSZ constructions which give the
actions \eqref{eq:AKSZAmod1} and \eqref{eq:AKSZAmod2}, and then relate
them to the topological A-model via suitable choices of Lagrangian
submanifolds $\mbf\cL\subset\mbf\cM$ (or gauge fixing). For this, we
equip $M_7=M_6\times S^1$ with a direct product metric, where $M_6$ is a six-dimensional
Riemannian manifold with $SU(3)$-structure, and write the $G_2$-structure on
$M_7$ as
\be
\Phi = \dd X^7\w B + \rho \ ,
\ee 
where $B$ is an almost K\"ahler form of type $(1,1)$ with respect to the almost complex structure defined by the three-form $\rho$ on $M_6$. If $B$ and $\rho$ are independent of $X^7$, then $\dd\Phi=0$ implies $\dd B=\dd\rho=0$ and $M_6$ is
a Calabi-Yau threefold, as in the A-model topological string theory,
where $\rho$ is the real part of the global
holomorphic three-form $\Omega$ on $M_6$. However, for the purposes of
our ensuing AKSZ constructions only the K\"ahler class of the Calabi-Yau structure
is required, as in~\cite{AKSZ1997}. In particular, double dimensional
reduction on a circle of the flux deformation $I_{\Sigma_3,\Phi}$ along the lines
of \S\ref{sec:DimRedflux} gives the $B$-field coupling
$I_{\Sigma_2,w\,B}$ for the topological string, whose AKSZ
construction is given by the Poisson sigma-model of~\S\ref{sec:AKSZPoisson}. Hence in what
follows we shall only require that $M_6$ be a K\"ahler manifold.

\subsection{The topological A-model}
\label{sec:AmodeAuxF}

Let us briefly review the topological A-model, whose Mathai-Quillen formalism is given in e.g.~\cite{MQWu1995}. It is defined by maps $X^i=(X^a,X^{\bar a})$ from the worldsheet $\Sigma_2$ to the K\"ahler manifold $M_6$, where $a=1,2,3$ are complex target space indices and we use local complex coordinates $\sigma=(z,\bar z)$ on the Riemann surface $\Sigma_2$. We further introduce ghost fields $(\chi^a_{\bar{z}},\chi^{\bar{a}}_z,\psi^a,\psi^{\bar{a}})$ with ghost number $(-1,-1,1,1)$. The action of the topological A-model is then
\be \label{eq:Amod}
I_{\rm{A}} \, = \, \int_{\Sigma_2}\, \dd^2 z\ \Big( \,g_{a\bar{b}} \, \partial_{\bar{z}}X^a\, \pa_z X^{\bar{b}} \, + \, \ii g_{a\bar{b}}\,\big( \chi^a_{\bar{z}}\, \nabla_z \psi^{\bar{b}} + \chi^{\bar{b}}_z\, \nabla_{\bar{z}} \psi^a\big)  \, - \, R_{a \bar{b} c \bar{d}} \, \chi^a_{\bar{z}}\, \chi^{\bar{b}}_z\, \psi^c\, \psi^{\bar{d}}\, \Big) \  ,
\ee
where $g_{a\bar{b}}$ is the K{\"a}hler metric which obeys the K{\"a}hler identity $\partial_a g_{b\bar{c}}=\partial_b g_{a\bar{c}}$ and its complex conjugate $\partial_{\bar{a}} g_{b\bar{c}}=\partial_{\bar{c}} g_{b\bar{a}}$. The Levi-Civita connection is $\nabla_z \psi^{\bar{a}}=\partial_{z}\psi^{\bar{a}} + \Gamma^{\bar{a}}{}_{\bar{b}\bar{c}}\, \psi^{\bar{b}}\, \pa_z X^{\bar{c}}$ and its complex conjugate $\nabla_{\bar{z}} \psi^{a}=\partial_{\bar{z}}\psi^{a} + \Gamma^{a}{}_{bc}\, \psi^{b}\, \pa_{\bar{z}} X^{c}$. The complex Christoffel symbols are ${\Gamma^a}_{bc}=g^{a\bar{d}}\, \Gamma_{\bar{d}bc}$ and ${\Gamma^{\bar{a}}}_{\bar{b}\bar{c}}=g^{d\bar{a}}\, \Gamma_{d\bar{b}\bar{c}}$, where
$\Gamma_{a\bar{b}\bar{c}}=\partial_{\bar{b}} g_{a\bar{c}}$ and $\Gamma_{\bar{a}bc}=\partial_b g_{c\bar{a}}$. The Riemann tensor is $R_{a \bar{b} c \bar{d}}= -g_{a\bar{e}}\, \partial_c {\Gamma^{\bar{e}}}_{\bar{b}\bar{d}}$. The action \eqref{eq:Amod} is invariant under the BRST transformations
\be\label{eq:BRSTAmod}
\delta X^{a} \,  = \, \ii \psi^{a}\  , \qquad \delta X^{\bar{a}} \,  = \, \ii \psi^{\bar{a}} \ , \qquad \delta \psi^{a} \,  = \, 0 \qquad \mbox{and} \qquad \delta \psi^{\bar{a}}\, = 0 \ , 
\ee
together with
\be 
\delta \chi^a_{\bar{z}} \,  = \,- \, \partial_{\bar{z}} X^{a} \, - \, \ii{\Gamma^a}_{bc}\,\psi^b\, \chi^c_{\bar{z}}  \qquad \mbox{and} \qquad
\delta\chi^{\bar{a}}_{z} \,  = \,- \, \partial_{z} X^{\bar{a}} \, - \, \ii{\Gamma^{\bar{a}}}_{\bar{b}\bar{k}}\, \psi^{\bar{b}}\, \chi^{\bar{c}}_{z}\ .
\ee  
The fixed point locus of the BRST charge is the space of holomorphic maps $X:\Sigma_2\to M_6$, which are worldsheet instantons.

Let us now reformulate the topological A-model with a linearizing auxiliary field, analogously to what we did in \S\ref{sec:MQmembrAuxF} for the Mathai-Quillen membrane sigma-model. We introduce two fields $b^a_{\zbar}$ and $b^{\bar{a}}_z$ with ghost number~0, and the new BRST transformations given by \eqref{eq:BRSTAmod} together with
\be 
\delta \chi^a_{\bar{z}} \,  =  \, b^a_{\zbar} \  , \qquad
\delta\chi^{\bar{a}}_{z} \,  = \, b^{\bar{a}}_z \  , \qquad
 \delta b^a_{\zbar} \,  =  \, 0  \qquad \mbox{and} \qquad
\delta b^{\bar{a}}_z \,  = \, 0 \  .
\ee  
The action
\be \label{eq:AmodBRSTeg}
S_{\rm{A}}  \, = \, \delta \Psi_{\rm{A}} \  ,
\ee
with the gauge fixing fermion 
\be \label{eq:GFFermionAmodel} \begin{aligned}
\Psi_{\rm{A}} \, =& \, - \int_{\Sigma_2}\, \dd^2 z\ \Big( g_{a\bar{b}}\,\big(\chi^a_{\bar{z}}\, \partial_z X^{\bar{b}} \, + \, \chi^{\bar{a}}_z\, \partial_{\bar{z}}X^b \big) \, + \, 
\frac 12\, g_{a\overline{b}}\, \big(\chi^a_{\overline{z}}\, b^{\overline{b}}_z \, + \, \chi^{\overline{a}}_{z}\, b^b_{\overline{z}}\big) \\
& \qquad \qquad \qquad \, + \, 
\frac{\ii}{2}\, \Gamma_{a\bar{b}\bar{c}}\, \psi^{\bar{b}}\, \chi^{\bar{c}}_z\, \chi^a_{\bar{z}} \, + \, \frac{\ii}{2}\,\Gamma_{\bar{a}bc}\, \psi^{b}\, \chi^{c}_{\bar{z}}\, \chi^{\bar{a}}_{z} \Big) \  ,
\end{aligned}\ee
reduces to the action \eqref{eq:Amod} after using the equations of motion of the auxiliary fields $b^a_{\zbar}$ and $b^{\bar{a}}_z$ which give
\be 
 b^a_{\bar{z}} \, = \,- \, \partial_{\bar{z}} X^{a} \, - \, \ii{\Gamma^a}_{bc}\, \psi^b\, \chi^c_{\bar{z}}  \qquad \mbox{and} \qquad
b^{\bar{a}}_{z} \, = \,- \, \partial_{z} X^{\bar{a}} \, - \, \ii{\Gamma^{\bar{a}}}_{\bar{b}\bar{c}}\, \psi^{\bar{b}}\, \chi^{\bar{c}}_{z}\  .
\ee  

\subsection{BV formulation and AKSZ constructions}
\label{sec:topAmodI}
  
If we define the antifields as 
\be 
X^+_{a} \,= \, \frac{\d\Psi_{\rm{A}}}{\delta X^{a}} \ , \qquad X^+_{\bar{a}} \,= \, \frac{\d\Psi_{\rm{A}}}{\delta X^{\bar{a}}}\  , \qquad \big(\chi^+_{a}\big)_{z} \, = \,\frac{\rdelta\Psi_{\rm{A}}}{\delta \chi^a_{\zbar}} \qquad \mbox{and} \qquad \big(\chi^+_{\bar{a}}\big)_{\zbar} \, = \,\frac{\rdelta\Psi_{\rm{A}}}{\delta \chi^{\bar{a}}_{z}} \  ,
\ee
we can rewrite \eqref{eq:AmodBRSTeg} as a BV-type action
\be \label{eq:AmodBVaction}
S_{\rm{A}} \, = \, \int_{\Sigma_2}\, \dd^2 z\ \Big(\psi^{a}\, X^+_{a} \, + \, \psi^{\bar{a}}\, X^+_{\bar{a}} \, + \, b^{a}_{\bar{z}}\, \big(\chi^+_{a}\big)_{z} \, + \, b^{\bar{a}}_{z}\, \big(\chi^+_{\bar{a}}\big)_{\bar{z}}\Big) \  .
\ee
In the following we give two new AKSZ constructions for the topological A-model, which each differ from the Poisson sigma-model. 

\medskip

{\underline{\sl AKSZ construction I.} \ } 
Our first AKSZ construction for the topological A-model is analogous to the first AKSZ membrane sigma-model in \S\ref{sec:AKSZmembr1}. The source dg-manifold is the superworldsheet $\cW= T [1]\Sigma_2$, while the target symplectic dg-manifold is $\cM= { T }^*[1] T [-1] T [1]M_6$, where $M_6$ is a K\"ahler manifold. The base coordinates in $ T [-1] T [1]M_6$ are $(X^i,q^i,b^i,\eta^i)$ with degree $(0,1,0,-1)$, where $X^i$ are associated to the coordinates of $M_6$. The graded fiber coordinates are $(\chi_i,p_i, n_i, h_i)$ with degree $(1,0,1,2)$. The canonical symplectic structure of degree~1 on the target superspace $\cM$ is
\be \label{eq:AKSZsymplAmod1}
\omega_{2, \rm{I}} \, = \, \dd \chi_i \w \dd X^i \, + \, \dd p_i \w \dd q^i \, + \, \dd h_i \w \dd \eta^i \, + \, \dd n_i \w \dd b^i  \ .
\ee
This gives the same BV symplectic structure on the space $\mbf\cM$ of superfields as in \eqref{eq:AKSZsymplAmodBV1}, and the AKSZ action is \eqref{eq:AKSZAmod1}.

We choose the Lagrangian submanifold $\mbf\cL\subset\mbf\cM$ given by
\be
\chi^{(0)} \,=\, \chi^{(1)} \,=\,0 \  , \qquad p^{(0)}\,= \,p^{(1)}\,= \, 0 \ , \qquad n^{(0)}\, = \,n^{(1)}\, = \, 0  \qquad \mbox{and} \qquad h^{(0)}\, = \,h^{(1)}\, = \, 0 \ .
\ee
Writing the coordinate indices of the K{\"a}hler manifold $M_6$ as before in complex notation $i=(a,\bar{a})$, where $a=1,2,3$, and the complex coordinates on the worldsheet $\Sigma_2$ as $(z,\bar{z})$, we define the component fields 
\be \begin{aligned}
X^{(0)\, i} & \,=\, X^{i}\  , & \qquad \chi^{(2)}_{i} & \,=\,  X^+_{i}\  , & \qquad q^{(0)\, i} & \,=\, \psi^{i} & \qquad &\mbox{and}& \quad p^{(2)}_{i} & \,=\, - \, \psi^+_{i}\  , \\[4pt]
b^{(0)\, a} & \,=\, b^{a}_{\bar{z}}\  , & \qquad b^{(0)\, \bar{a}} & \,=\, b^{\bar{a}}_{z}\  , & \qquad {n}^{(2)}_{a} & \,=\, \big(b^{+}_{a}\big)_{z}  & \qquad &\mbox{and}& \quad {n}^{(2)}_{\bar{a}} & \,=\,  \big(b^{+}_{\bar{a}}\big)_{\bar{z}}\  ,  \\[4pt]
\eta^{(0)\, a} & \,= \, \chi^{a}_{\bar{z}}\  , & \qquad \chi^{(0)\, \bar{a}} & \,=\, \chi^{\bar{a}}_{z}\  , & \qquad h^{(2)}_{a} & \,= \, - \, \big(\chi^{+}_{a}\big)_{z}  & \qquad &\mbox{and} \quad & h^{(2)}_{\bar{a}} & \,=\, - \, \big(\chi^{+}_{\bar{a}}\big)_{\bar{z}}\  . 
\end{aligned} \ee
With this notation, the restriction of the AKSZ action \eqref{eq:AKSZAmod1} to $\mbf\cL$ yields the BV action \eqref{eq:AmodBVaction} as $\bs{ S}_{\rm{A}, \rm{I}}\big|_{\mbf\cL}= -S_{\rm{A}}$, and it can be gauge fixed further to the A-model action with gauge fixing fermion $-\Psi_{\rm{A}}$ from \eqref{eq:GFFermionAmodel}.

\medskip

{\underline{\sl AKSZ construction II.} \ } 
We introduce a second AKSZ construction for the topological A-model, which is the analogue of the second AKSZ membrane sigma-model in \S\ref{sec:AKSZmembr1}. We start with the same source dg-manifold $\cW= T [1]\Sigma_2$ as in the previous construction, but now we choose $\cM= T ^*[1] T [1]M_6$ to be the target QP-manifold of degree~1 with coordinates $(X^i,\chi_i,q^i,p_i)$ with degree $(0,1,1,0)$. The symplectic structure
\be
\omega_{2, \rm{II}} \, = \, \dd \chi_i \w \dd X^i \, + \, \dd p_i \w \dd q^i 
\ee
is the restriction of \eqref{eq:AKSZsymplAmod1}. The AKSZ action is also the restriction \eqref{eq:AKSZAmod2}.

We introduce the component fields
\be\begin{aligned} 
X^{(0)\,i} & \, = \, X^{i} \  ,  & \qquad  \big(\chi^{(2)}_{i}\big)_{ z\zbar} & \, = \,   X^+_{i}\  , & \qquad q^{(0)\,i} & \, = \, \psi^{i}  &  \qquad &\mbox{and}& \qquad \big(q^{(2)}_{i}\big)_{ z\zbar} & \, = \,   \psi^+_{i} \  , \\[4pt]
X^{(1)\,a}_{\zbar} & \, = \, \chi^{a}_{\zbar}\  ,  &   \qquad \big(\chi^{(1)}_{a}\big)_{z}&\, = \, - \, \big(\chi^+_{a}\big)_{z} \  , & \qquad 
X^{(1)\,\bar{a}}_{z} & \, = \, \chi^{\bar{a}}_{z}  &   \qquad &\mbox{and}& \big(\chi^{(1)}_{\bar{a}}\big)_{\zbar}&\, = \, \big(\chi^+_{\bar{a}}\big)_{\zbar} \  , \\[4pt]
q^{(1)\,a}_{\zbar} & \, = \, b^{a}_{\zbar}\  ,  & \qquad \big(p^{(1)}_{a}\big)_{z}&\, = \, - \, \big(b^+_{i}\big)_{z}\  , & \qquad 
q^{(1)\,\bar{a}}_{z} & \, = \, b^{\bar{a}}_{z}  & \qquad &\mbox{and}& \big(p^{(1)}_{\bar{a}}\big)_{\zbar}&\, = \, \big(b^+_{\bar{a}}\big)_{\zbar} \  ,
\end{aligned} \ee
and choose the Lagrangian submanifold defined by
\be
\chi^{(0)\, i} \, = \, 0 \  , \qquad p^{(0)}_i \, = \, 0 \  ,  \qquad \big(\chi^{(1)}_{a}\big)_{\zbar} \, = \, \big(\chi^{(1)}_{\bar{a}}\big)_{z} \, = \, 0   \qquad \mbox{and} \qquad \big(p^{(1)}_{a}\big)_{\zbar} \, = \, \big(p^{(1)}_{\bar{a}}\big)_{z} \, = \, 0  \ .
\ee
This yields the same BV action $-S_{\rm{A}}$ from \eqref{eq:AmodBVaction}, which gives the A-model action with gauge fixing fermion $-\Psi_{\rm{A}}$ from \eqref{eq:GFFermionAmodel}. Note that in neither of these AKSZ constructions does the target dg-manifold coincide with that of the Poisson sigma-model from \S\ref{sec:AKSZPoisson} associated to string fields $X:\Sigma_2\to M_6$.

\subsection{Dimensional reduction from the standard Courant sigma-model}

The first AKSZ construction of the A-model in \S\ref{sec:topAmodI} can be embedded into the standard Courant sigma-model, which is a membrane theory, in a similar way as we embedded our AKSZ membrane sigma-models into the standard 2-Courant sigma-model, which is a threebrane theory, in \S\ref{sec:3braneAndMembrane}. 
For this, let us consider the standard Courant sigma-model from \S\ref{sec:CourantAlgAKSZ} on a product worldvolume $\Sigma_3=\Sigma_2\times S^1$, and assume that our superfields do not depend on the extra coordinate of $S^1$. The AKSZ action of the standard standard Courant sigma-model is given by \eqref{eq:AKSZactionStdCourant} 
and the BV symplectic form by \eqref{eq:StdCourantBVomega}. After integration over the extra supercoordinates on $T[1]S^1$ and a relabelling of superfields, we arrive at the AKSZ action
\be
\bs{S}_{2,\rm{red}} \, = \, \int_{ T [1]\Sigma_2}\,\dd^2\hat z \ \big({\bs{\chi}}_i\, {\bs{q}}^i \, + \, {\bs{b}}^i\, {\bs{h}}_i \, - \, \bs{\chi}_i\, \bs{D} \bs{X}^i \, - \, \bs{h}_i\, \bs{D} \bs{\eta}^i \, - \, \bs{n}_i\, \bs{D} \bs{b}^i \, - \, \bs{p}_i\, \bs{D} \bs{q}^i \big) \  ,
\ee
and the BV symplectic form 
\be 
\bs{\omega}_{2,\rm{red}} \, = \, \int_{ T [1]\Sigma_2}\,\dd^2\hat z \ \big(\bs{\delta} \bs{\chi}_i\, \bs{\delta} \bs{X}^i \, + \, \bs{\delta} \bs{p}_i\, \bs{\delta} \bs{q}^i \, + \, \bs{\delta} \bs{h}_i\, \bs{\delta} \bs{\eta}^i \, + \, \bs{\delta} {\bs{n}}_i\, \bs{\delta} \bs{b}^i \big) \ .
\ee
This symplectic form is the same as that of the A-model in \eqref{eq:AKSZsymplAmodBV1}, and the AKSZ action reduces to the A-model action \eqref{eq:AKSZAmod1} if we set the kinetic terms to zero by definition or via gauge fixing.

\section{AKSZ theory for supersymmetric quantum mechanics}
\label{sec:SQM}

In this section we continue the dimensional reduction procedure one final time, and reduce our second AKSZ construction of the A-model to an AKSZ formulation for supersymmetric quantum mechanics.
We have seen in \S\ref{sec:AKSZmembr1} and \S\ref{sec:topAmodI} that both the topological A-model and the topological membrane sigma-models on $G_2$-manifolds have similar AKSZ constructions. Following the same procedure as before we give an analogous AKSZ construction for supersymmetric quantum mechanics.

\subsection{Dimensional reduction of the A-model}

We start with the canonically transformed action from \eqref{eq:AmodDRAction} restricted to the fields of the second AKSZ construction:
\be 
\bs{ S}_{G_2,\rm{II}}^{\rm{eff}} \, = \, \int_{ T
  [1]\Sigma_2}\,\dd^2\hat z \ \Big({\bs{\chi}}_i\, {\bs{q}}^i \, + \, \varepsilon\, \big({\bs{\chi}}_i\, \bs{D} {\bs{X}}^i \, + \, {\bs{q}}^i\, \bs{D} {\bs{p}}_i \big) \Big) \ ,
\ee
and the corresponding symplectic structure from \eqref{eq:AKSZsymplAmodBV1}:
\be 
\bs{\omega}_{2,\rm{II}} \, = \, \int_{ T [1]\Sigma_2}\,\dd^2\hat z \ \big(\bs{\delta} \bs{\chi}_i\, \bs{\delta} \bs{X}^i \, + \, \bs{\delta} \bs{p}_i\, \bs{\delta} \bs{q}^i \big) \  .
\ee
We apply Losev's trick from \S\ref{sec:DimRedMeth} and use the same notation as in \eqref{eq:DimRedNotation} to calculate the reduction on a product source space $\Sigma_2=S^1 \times \Sigma_1$, where we distinguish the circle $\Sigma_1=S^1$ along which the dimensional reduction takes place. We choose $\wt{\bs{\chi}}_i$, $\wt{\bs{q}}{}^{\,i}$, $\bs{X}_t^i$ and $(\bs{p}_{t})_{i}$ to be the ultraviolet fields, and we set the gauge $\bs{X}_t^i=0$ and $\wt{\bs{q}}{}^{\,i}=0$. After integrating out the ultraviolet fields, we obtain the effective action 
\be 
\bs{ S}_{\mathrm{A},\rm{II}}^{\rm{eff}} \, = \, \int_{ T
  [1]S^1}\,\dd\hat z \ \Big({\bs{B}}_i\, {\bs{\xi}}^i \, + \, \varepsilon\, \big(-{\bs{B}}_i\, \bs{D} {\bs{X}}^i \, - \, {\bs{\xi}}^i\, \bs{D}\, {\bs{\eta}}_i \big) \Big) \  ,
\ee	
and the symplectic structure
\be \label{eq:AKSZSymplSQM}
\bs{\omega}_{1} \, = \, \int_{ T [1]S^1}\,\dd\hat z \ \big(\bs{\delta} \bs{B}_i\, \bs{\delta} \bs{X}^i \, + \, \bs{\delta} \bs{\eta}_i\, \bs{\delta} \bs{\xi}^i \big) \  ,
\ee	
where we relabeled the fields as
\be 
{\bs{B}}_i \, = \, - \int_{\Sigma_1}\, \dd t \  ({\bs{\chi}}_{t})_{i} \  , \qquad  {\bs{X}}_i \, = \, \wt{\bs{X}}{}^i \  , \qquad  {\bs{\eta}}_i \, = \, - \, \int_{\Sigma_1}\, \dd t \ ({\bs{p}}_{t})_{i} \qquad \mbox{and} \qquad \bs{\xi}^i \, = \, - \, \wt{\bs{q}}{}^{\,i} \  , 
\ee
and these new fields are independent of the coordinate $t$ of $\Sigma_1$.

The infinitesimal canonical transformation \eqref{eq:inftezCanTrafo} with the fermionic functional
\be
\bs{\alpha} \, = \, \int_{ T [1]S^1}\,\dd\hat z \ \bs{\eta}_i\, \bs{D} \bs{X}^i 
\ee
gives the action
\be \label{eq:AKSZSSQM}
\bs{S}_{\rm{SQM}} \, = \, \int_{ T [1]S^1}\,\dd\hat z \ \bs{B}_i\, \bs{\xi}^i \, .
\ee
We will see in \S\ref{sec:AKSZSQM} below that this action gives an AKSZ formulation of supersymmetric quantum mechanics. Nothing we discuss in this section depends on the target space K\"ahler structure nor even on its dimensionality, and the reduction of the topological sigma-model described here applies to generic maps whose target is any Riemannian manifold.

\subsection{Supersymmetric quantum mechanics}

Supersymmetric quantum mechanics provides a simple example of a topological field theory; its Mathai-Quillen formalism can be found in e.g.~\cite{MQWu1995,MQAni2005,MQBlau1995}. The target space is a Riemannian manifold $M$ with metric $g$ and the parameter manifold is just a compact worldline $S^1$. The local coordinates of the mapping space $LM:=\mathsf{Map}(S^1,M)$ are $\{x^i(\tau)\}$ with $\tau\in[0,1]$ and $x^i(0)=x^i(1)$, and so they parameterize (smooth) loops in $M$. We furthermore define two fermionic fields $\psi^i(\tau)$ and $\bar{\psi}_i(\tau)$ with ghost number $1$ and $-1$, respectively. The action of supersymmetric quantum mechanics is
\be \label{eq:ISQM}
I_{\mathrm{SQM}} \, = \, \int_{S^1}\, \dd \tau\ \Big(\,\frac 12\, g_{ij}\, \dot{x}{}^i\, \dot{x}{}^j \, + \, \ii \bar{\psi}_i\, \nabla_\tau \psi^i \, - \, \frac 14\, R^{ij}{}_{kl}\, \bar\psi_i\, \bar\psi_j\, \psi^k\, \psi^l\, \Big) \  ,
\ee
where a dot denotes a $\tau$-derivative, $\nabla_\tau \psi^i = \dot{\psi}{}^i + {\Gamma^i}_{jk}\, \psi^j\, \dot{x}{}^k$ is defined by the action of the Levi-Civita connection $\nabla$ of the metric $g$ pulled back to the loop via the map $x$, and $R$ is the associated Riemann tensor. 
The action \eqref{eq:ISQM} is invariant under the BRST transformations
\be
\delta x^i \, = \, \psi^i \ , \qquad \delta \psi^i \, = \, 0 \qquad \mbox{and} \qquad \delta \bar{\psi}_i \, = \, \ii g_{ij}\, \dot{x}^j \, + \, {\Gamma^k}_{ij}\, \psi^j\, \bar{\psi}_k \  ,
\ee
which is only nilpotent on-shell, and it is BRST-exact on-shell:
\be \label{eq:ISQMpotential}
I_{\mathrm{SQM}} \, = \, \delta \Psi'_{\mathrm{SQM}} \qquad \mbox{with} \quad \Psi'_{\mathrm{SQM}} \, = \, - \frac{\ii}{2}\, \int_{S^1}\, \dd\tau \ \bar{\psi}_i\, \dot{x}{}^i \ . 
\ee
The set of $\delta$-fixed points is the space of instantons, i.e. the constant loops $x^i(\tau)$, which can be identified with the target space $M$.

We follow the same procedure as in \S\ref{sec:MQmembrAuxF} and
\S\ref{sec:AmodeAuxF} to reformulate supersymmetric quantum mechanics
using a linearizing auxiliary field $b_i$ with ghost number~0. The
BRST transformations with the field~$b_i$ are given by
\be
\delta x^i \, = \, \psi^i \ , \qquad \delta \psi^i \, = \, 0  \ ,
\qquad \delta \bar{\psi}_i \, = \, b_i \qquad \mbox{and} \qquad \delta
b_i \, = \, 0 \ ,
\ee
and they are nilpotent off-shell. The action
\be
\delta\Psi'_{\mathrm{SQM}} \, = \, - \frac{\ii}{2}\, \int_{S^1}\, \dd\tau \  \big(  b_i\, \dot{x}{}^i \, - \, \bar{\psi}_i\, \dot{\psi}{}^i  \big) 
\ee
is invariant under these new BRST transformations, and it reduces to the action \eqref{eq:ISQM} if we impose the constraint
\be \label{eq:b_i_constraint}
b_i \, = \, \ii g_{ij}\, \dot{x}^j \, - \, {\Gamma^k}_{ij}\, \bar{\psi}_k\, \psi^j
\ee
as gauge fixing. In the language of the BRST formulation, this means that we choose the gauge fixing fermion as 
\be \label{eq:PsiprimeGF}
{\Psi}_{\mathrm{SQM}} \, = \, - \int_{S^1}\, \dd\tau \  \bar{\psi}_i\, \Big( \ii \dot{x}{}^i \, + \, \frac 12\, g^{jl}\,{\Gamma^i}_{lk}\, \bar{\psi}_j\, \psi^k \, - \, \frac 12\, g^{ij}\,b_j   \Big) \  .
\ee
The BRST variation of \eqref{eq:PsiprimeGF} gives us the action
\be \begin{aligned} \label{eq:S_SQMprime}
{S}_{\mathrm{SQM}} \, = \, \delta{\Psi}_{\mathrm{SQM}} \, = \,   \int_{S^1}\, \dd \tau \  \Big( & \ii \bar{\psi}_i\, \dot{\psi}{}^i \, - \, \ii b_i\, \dot{x}{}^i \, + \,  g^{il}\,{\Gamma^j}_{kl}\,\bar{\psi}_j\,\psi^k\, b_i \, + \, \frac 12\, g^{ij}\, b_i\, b_j  \\
&  - \, \frac 12\, \partial_k\big(g^{jm}\,{\Gamma^i}_{ml}\big)\,\bar{\psi}_i\, \bar{\psi}_j\, \psi^k\, \psi^l       \Big) \ .
\end{aligned}\ee
The equation of motion for $b_i$ gives the same field redefinition as in \eqref{eq:b_i_constraint}, and using this we find that the action \eqref{eq:S_SQMprime} is classically equivalent to the action \eqref{eq:ISQM}.

\subsection{AKSZ construction}
\label{sec:AKSZSQM}

Following the procedure in \S\ref{sec:AKSZmembr1} and \S\ref{sec:topAmodI}, we give an AKSZ formulation of supersymmetric quantum mechanics which reduces to the action \eqref{eq:ISQM} after gauge fixing and eliminating the auxiliary field $b_i$. 
Our source dg-manifold is $\cW= T [1]S^1$ and the target symplectic dg-manifold is $\cM={ T }^*( T [1]M)$, where $M$ is a Riemannian manifold with metric $g$. Denote the degree~0 and~1 coordinates of $ T [1]M$ by $X^i$ and $\xi^i$, respectively, and their cotangent coordinates by $B_i$ and $\eta_i$ with degree~0 and~$-1$, respectively. The canonical symplectic structure on $\cM={ T }^*( T [1]M)$ is
\be
\omega_1 \, = \, \dd B_i \w \dd X^i \, + \, \dd \eta_i \w \dd \xi^i \  ,
\ee
which gives the same symplectic form on the mapping space of superfields $\mbf\cM$ as in \eqref{eq:AKSZSymplSQM}. The AKSZ superfields are expanded as
\be \begin{aligned}
\bs{X}^i & \,=\,   x^i \, - \, b^{+ \, i}\, \theta \  , \\[4pt]
\bs{B}_i & \,=\,  - \, b_i \, + \, x^+_i\, \theta \  ,\\[4pt]
\bs{\xi}^i & \,=\,  - \, \psi^i \, + \, \bar{\psi}^{+ \, i}\, \theta \  ,\\[4pt]
\bs{\eta}_i & \,=\,  \bar{\psi}_i \, - \, \psi^+_i\, \theta \  ,
\end{aligned} \ee 
where the superworldline coordinate $\theta$ has degree 1. Our choice for the AKSZ action is the same as that in \eqref{eq:AKSZSSQM} which was obtained from the dimensional reduction of the A-model:
\be
\bs{S}_{\mathrm{SQM}} \, = \, \int_{ T [1]S^1}\,\dd\hat z \ \bs{B}_i\, \bs{\xi}^i  \, = \, - \int_{S^1}\, \dd\tau\ \big(   \psi^i\, x^+_i \, + b_i\,  \bar{\psi}^{+ \, i}  \big) \  ,
\ee
and it trivially solves the classical master equation $(
\bs{S}_{\mathrm{SQM}}, \bs{S}_{\mathrm{SQM}} )_{\mathrm{BV}}=0$. The
BV--BRST transformations\footnote{The action $\bs{S}_{\mathrm{SQM}}$
  is also invariant under the transformations $\delta x^+_i= b_i$,
  $\delta \bar{\psi}^{+\, i} = \psi^i$, $\delta \psi^i=0$ and $\delta
  b_i = 0$, and under the transformations $\delta x^+_i= 0$, $\delta
  \bar{\psi}^{+\, i} = 0$, $\delta \psi^i=\bar{\psi}^{+\, i}$ and
  $\delta b_i = -x^+_i$, but our transformations do not include
  these.} are generated by the cohomological vector field given by the
BV bracket $\mbf Q_{\rm{SQM}} = ( \bs{S}_{\mathrm{SQM}}, \, \cdot \, )_{\mathrm{BV}}$ and read as
\be\begin{aligned}
\mbf Q_{\rm{SQM}} x^i  &\,= \, \psi^i &\qquad &\mbox{and}& \qquad \mbf Q_{\rm{SQM}}  \psi^+_i & \,=\, x^+_i \ , \\[4pt]
\mbf Q_{\rm{SQM}}  \psi^i  &\,=\, 0 &\qquad &\mbox{and}& \qquad \mbf Q_{\rm{SQM}} x^+_i  &\,=\, 0 \ , \\[4pt]
\mbf Q_{\rm{SQM}}  \bar{\psi}_i  &\,=\,  b_i &\qquad &\mbox{and}& \qquad \mbf Q_{\rm{SQM}}  b^{+\, i}&\,=\, - \, \bar{\psi}^{+\, i} \ , \\[4pt]
\mbf Q_{\rm{SQM}}  b_i  &\,=\,  0 &\qquad &\mbox{and}& \qquad \mbf Q_{\rm{SQM}}\bar{\psi}^{+ \, i} & \,=\, 0 \  .
\end{aligned}\ee
The nilpotent fermionic symmetry $\mbf Q_{\rm{SQM}} $ acts trivially on the AKSZ action $\bs{S}_{\mathrm{SQM}}$.

We reduce the action $\bs{S}_{\mathrm{SQM}}$ to $I_{\mathrm{SQM}}$ after gauge fixing. We choose the same gauge fixing fermion $-{\Psi}_{\mathrm{SQM}}$ as in \eqref{eq:PsiprimeGF}. The pertinent antifields are given by 
\be\begin{aligned}
  x^+_i  &= \, - \, \frac{\delta{\Psi}_{\mathrm{SQM}}}{\delta x^i}  =  - \ii \dot{\bar{\psi}}_i \, + \, \frac 12\, \partial_i \big( {\Gamma^j}_{ml}\,g^{mk} \big)\,  \bar{\psi}_j\,  \bar{\psi}_k\, \psi^l \, - \, \frac 12\, \partial_i g^{jk}\, \bar{\psi}_j\, b_k \  , \\[4pt]
 \bar{\psi}^{+ \, i} &= \, - \, \frac{\rdelta{\Psi}_{\mathrm{SQM}}}{\delta \bar{\psi}_i } = \ii \dot{x}{}^i \, + \, {\Gamma^{[ i }}_{lk}\, g^{ j] l}\, \bar{\psi}_j\, \psi^k \, - \, \frac 12\, g^{ij}\, b_j \ ,
\end{aligned}\ee
where the other gauge fixing equations are not important here. Calculating the gauge fixed action of $\bs{S}_{\mathrm{SQM}}$ we get the action \eqref{eq:S_SQMprime}, which is classically equivalent to $I_{\mathrm{SQM}}$.
	
\section{Conclusions and outlook}
\label{sec:Conc}

In this paper we have constructed BV quantized topological membrane theories on $G_2$-manifolds using the AKSZ formulation, which unify the topological membrane theories of~\cite{MQAni2005} and~\cite{Bonelli2005b}. We have dimensionally reduced them to the A-model, and one of them has been reduced further to supersymmetric quantum mechanics. We also studied the derived bracket of one of our AKSZ topological membrane theories whose target is a derived symplectic dg-manifold with fields of negative degree, which gave an $L_\infty$-extension of the standard Courant bracket. It would be interesting to study further the consequences of this more complex derived algebroid structure.

We have further proposed a topological threebrane model given by the AKSZ construction, which reduces to our AKSZ membrane theories upon worldvolume dimensional reduction. Its derived bracket is the standard 2-Courant bracket, which appears in exceptional generalized geometry as the antisymmetrization of the generalized Lie derivative, and it is also the induced bracket of anomaly-free current algebras of topological membranes on $G_2$-manifolds~\cite{Bonelli2005a}. We have found that double dimensional reduction on a circle of our threebrane model with $G$-flux twisting yields the twisted standard Courant sigma-model, which geometrizes the $H$-flux in type~II string theory. 

Our constructions are the starting point for the introduction of exceptional generalized geometry~\cite{Berman2010,Berman2011,Berman2011b} and M-theory fluxes~\cite{Blair2014,Lust2017} for membranes in M-theory described by the AKSZ formalism. The first step towards this goal is our AKSZ threebrane sigma-model with its derived standard 2-Courant bracket. However, implementing non-geometric M-theory fluxes into this setting seems somewhat perplexing. In the string theory setting, 
T-duality in AKSZ membrane theory acts as a duality between standard and contravariant Courant sigma-models, and also transforms geometric $H$-flux and non-geometric $R$-flux into each other~\cite{Bessho2015}. It is tempting to try lifting this T-duality to a U-duality at the level of AKSZ threebrane theory, which transforms our threebrane into another topological threebrane with non-geometric flux. In the case of the Courant sigma-models, the duality interchanges the degree~1 coordinates $\psi^i$ and $\chi_i$, and it is implemented as a canonical transformation given by a bivector and its T-dual two-form $B$-field. For the 2-Courant sigma-models, it is natural to expect that there similarly exist canonical transformations which implement the interchange between the degree~2 quantities $\psi^i\, \psi^j$ and $\chi_i$. In this case a trivector and a three-form would arise, which should be related to the trivector and three-form $C$-field in exceptional generalized geometry. But unfortunately this does not seem to be the case as there are no symplectomorphisms which interchange $\psi^i\, \psi^j$ and $\chi_i$. Thus implementing U-duality and non-geometric fluxes seems to be far more complicated than in the string theory case.

We close by discussing some open avenues for future investigation. In~\cite{deBoer2006} a closed string on a $G_2$-manifold has been proposed as the dual of a topological $G_2$ membrane, and its quantization at one-loop order is considered in~\cite{deBoer2007}, which may be relevant to the quantization of our membrane construction that is of interest when considering its connection to physical string theory (see also~\cite{MQAni2005}). Likewise an open $G_2$ string theory is introduced in~\cite{deBoer2006b}, wherein the worldvolume theory of associative three-cycles has a membrane formulation given by a gauge fixed Chern-Simons theory coupled to normal deformations of the cycle. A further development would be to give an AKSZ construction for this three-cycle theory, and to compare it with our AKSZ topological membrane theories. It would also be interesting to study the topological membrane of~\cite{Bao2005} in the context of the AKSZ construction. Finally, in the present paper we also derived AKSZ constructions for the A-model, hence one of the applications of our results is to study the possible dualities between the A-model and the B-model at the level of the AKSZ formalism, and in particular to find a realization of S-duality~\cite{Nekrasov2004b} in AKSZ theory. In this respect it would be interesting to study further the threebrane theory of calibrated four-cycles on eight-dimensional $Spin(7)$-manifolds that we discussed in Section~\ref{sec:3braneAndMembrane}, which may be relevant to the study of S-duality as in~\cite{Anguelova2004}.

\section*{Acknowledgments}

We thank Alberto Cattaneo, Andreas Deser, Branislav Jur\v{c}o, Emanuel Malek, Christian S\"amann, Peter Schupp and Satoshi Watamura for helpful discussions.
This work was supported by the Action MP1405 QSPACE, funded by the European Cooperation in Science and Technology (COST).
The work of Z.K. was supported by the New National Excellence Program of the Hungarian Ministry of Human Capacities and by the Hungarian Research Fund (OTKA). The work of R.J.S. was
supported by the Consolidated Grant ST/P000363/1 
from the UK Science and Technology Facilities Council.

\end{document}